\documentclass{iopart}
\expandafter\let\csname equation*\endcsname\relax
\expandafter\let\csname endequation*\endcsname\relax
\usepackage{ulem}
\usepackage{graphicx,color}
\usepackage{amsmath}
\usepackage{amssymb}
\usepackage[colorlinks=true,citecolor=blue]{hyperref}
\usepackage{cite}

\newcommand{\Up}{{\uparrow}}
\newcommand{\Dn}{{\downarrow}}
\newcommand\bwt         {\begin{widetext}}
\newcommand\ewt         {\end{widetext}}

\newcommand{\beq}{\begin{equation}}
\newcommand{\eeq}{\end{equation}}

\begin{document}

\title[Exact diagonalization of lattice models in Abelian and non-Abelian potentials]{Exact diagonalization of cubic lattice models in commensurate Abelian magnetic fluxes and translational invariant non-Abelian potentials}

\author{M. Burrello$^1$, I.C. Fulga$^2$, L. Lepori$^{3,4}$, A. Trombettoni$^{5,6}$}

\address{$^1$Niels Bohr International Academy and Center for Quantum Devices, NBI, University of Copenhagen, Juliane Maries Vej 30,
2100 Copenhagen, Denmark.}

\address{$^2$Institute for Theoretical Solid State Physics, IFW Dresden, 01171 Dresden, Germany.}

\address{$^3$Dipartimento di Scienze Fisiche e Chimiche, Universit\'a dell'Aquila, via Vetoio, I-67010 Coppito-L'Aquila, Italy.}
\address{$^4$INFN, Laboratori Nazionali del Gran Sasso, Via G. Acitelli, 22, I-67100 Assergi (AQ), Italy.}

\address{$^5$SISSA and INFN, Sezione di Trieste,
via Bonomea 265, I-34136 Trieste, Italy.}
\address{$^6$CNR-IOM DEMOCRITOS Simulation Center, Via Bonomea 265, I-34136 Trieste, Italy.}
\ead{michele.burrello@nbi.ku.dk}

\begin{abstract}
We present a general analytical formalism to determine the energy spectrum of 
a quantum particle in a cubic lattice 
subject to translationally invariant commensurate magnetic fluxes and 
in the presence of a general space-independent non-Abelian gauge potential. 
We first review and analyze the case of purely Abelian potentials, showing also that
the so-called Hasegawa gauge yields a decomposition of the Hamiltonian into sub-matrices having 
minimal dimension. Explicit expressions for such matrices are derived, also for general 
anisotropic fluxes. Later on, we show that the introduction of a translational invariant non-Abelian coupling for multi-component spinors does not affect the dimension of the minimal Hamiltonian blocks, 
nor the dimension of the magnetic Brillouin zone. General formulas are presented for the $U(2)$ case and
explicit examples are investigated involving $\pi$ and $2\pi/3$ magnetic fluxes. 
Finally, we numerically study the effect of random flux perturbations. 
\end{abstract}

\maketitle

\section{Introduction}
\label{sec:intro}

The study of the effects of a magnetic field on a 
quantum particle and on its energy spectrum 
is a subject of research as old as quantum mechanics, with a plethora of 
applications ranging from the Aharonov-Bohm effect \cite{aharonov59} to 
spintronics \cite{zutic04} and the quantum Hall effect 
\cite{yoshioka}. A special role in this field is certainly played by the study 
of properties of particles subject to the combined effect of a periodic 
potential and a magnetic field, 
starting from the classic papers on the energy spectrum 
of Bloch electrons in rational and irrational magnetic 
fields \cite{azbel64,hofstadter76}. 
The durable interest for the study of the interplay 
between the discreteness introduced by the 
lattice potential and the effects of the magnetic fields 
is motivated both by the physical relevance and by the
mathematical beauty of these systems and their variants, 
including the relation with the one-dimensional Harper model 
\cite{harper55,thouless90}, 
incommensurability effects \cite{hofstadter76,aubry} and topological invariants 
\cite{tkkn}. 

The possibility of controlling the intensity and the properties 
of the applied magnetic fields play a crucial role and  provides an essential 
tool to explore a rich variety of phenomena, occurring already 
at the single particle level, as clear 
for the Hofstadter problem \cite{hofstadter76}. 
Since gauge potentials can be exploited to modify and control the particle 
dynamics, they also 
provide an instrument to break or create new symmetries 
and to engineer non-trivial band structures, 
as exemplified by the integer quantum Hall effect, obtained just from the application of the simplest gauge potential, 
a constant magnetic field. 
When the dynamics of the electrons or atoms is coupled to some 
inner degree of freedom, 
as in the case of the spin-orbit coupling, then the minimal coupling can be done 
with non-Abelian gauge potentials. The non-trivial combination of pseudospin degrees of freedom and a non-Abelian lattice dynamics is indeed crucial for 
the implementation of many important models discussed in 
the last decade, 
including topological insulators and superconductors 
\cite{zhang11,hasankane,ludwig08}.

Beside the successes in realizing such models in solid state devices, 
the recent experimental developments in the field of 
ultracold atomic gases \cite{bloch08,LibroAnna,dalibard11} 
opened new scenarios to realize Abelian and non-Abelian gauge 
potentials in optical lattices, for instance imposing 
laser-assisted tunneling amplitudes to trapped atoms 
\cite{dalibard11,goldmanrev} (see as well the recent review \cite{chapter} 
in the book \cite{libro}). The possibility of implementing 
tunable gauge potentials became a paradigmatic 
example of the tools that can be exploited 
in the experimental design of novel quantum phases of matter and 
is providing a remarkable arena of challenging mathematical developments.

Such tools prompted a huge variety of theoretical investigations, 
aimed to propose new realizations of non-trivial phenomena and models, 
including the study of the physics of the 
Hofstadter butterfly \cite{hofstadter76,jak2003,bloch2013}, 
Weyl, Dirac and Majorana fermions \cite{toolbox,LMT,lan11,tarruell2012,lepori2017}, 
extra dimensions \cite{extra,fallani2015,spielman2015} 
and the implementation of states 
with non-Abelian excitations \cite{burrello10}.  
As an example of application of synthetic gauge potentials, 
relevant for the purposes 
of this paper, we observe that in two dimensions one can 
obtain Dirac cones in square lattices with a magnetic $\pi$-flux 
(half of the elementary flux) 
threading each plaquette \cite{Juzeliunas,lim08,hou09,affleck}. 
Similar properties may also be obtained in three dimensions: 
cubic lattices with synthetic $\pi$ fluxes in each plaquette still allow
to obtain Weyl fermions \cite{LMT,Hasegawa,Zou,MLT}. 
Moreover, artificial non-Abelian potentials may enable 
the possibility to explore ranges of Hamiltonian parameters, for instance for the spin-orbit 
couplings, that would be difficult to achieve in corresponding 
solid state devices. Further advantages offered by these setups are 
the possibility to control the contact interactions 
through Feshbach resonances and to tune
independently the synthetic magnetic field 
and the Zeeman terms. 

The goal of this work is to provide a unified formalism to determine the 
energy spectrum of a quantum particle on a cubic lattice 
subject to translational invariant commensurate magnetic fluxes and
in the presence of a general  non-Abelian gauge potential, also position-independent. 
The reasons for such a study are twofold: {\textit{i)}} 
in most of the proposals and the experimental realizations listed 
above, the gauge potentials are translational invariant and, despite several 
interesting instances have been considered, 
we think that it is still useful  a systematic study of non-Abelian 
gauge configurations in the simultaneous presence of an Abelian magnetic field. 
{\textit{ii)}} the interplay of Abelian and non-Abelian gauge potentials poses in general 
fascinating mathematical questions. For instance,  
we show that the magnetic Brillouin zone defined in absence of the 
non-Abelian terms remains unaltered when a translational invariant one is added. We also show 
that for a commensurate Abelian potential  the optimal gauge choice
decomposing the Hamiltonian into matrices with
minimal dimension is the so-called Hasegawa gauge \cite{Hasegawa}.

After the general discussion about  non-Abelian gauge potentials, we examine the case of $U(2)$ 
gauge potentials, which is relevant for most of the  
realizations/studies mentioned above. We 
observe in Section \ref{general} that, while for an Abelian gauge configuration translational 
invariance is explicit and amounts  to have a homogeneous  magnetic field,
in the presence of non-Abelian potentials a different gauge invariant 
definition of translational invariance is required. Even though the non-Abelian configurations 
we focus on transform under $U(2)$, from the discussions in the text 
it will be clear that most of our results are still valid also 
for translationally invariant gauge configurations related to larger 
non-Abelian groups.  

The plan of the paper is the following. We recall in Section \ref{Abelian} 
the case of Abelian translational invariant configurations 
with commensurate flux $\Phi$, focusing on the definition and on the structure 
of the magnetic Brillouin zone (MBZ). For the sake to maintain the paper 
self-consistent, we discuss in detail the purely Abelian case, 
showing that the so-called Hasegawa gauge yields, for any commensurate magnetic flux, 
the minimal dimension of the Hamiltonian blocks in which the 
lattice Hamiltonian can be decomposed. For completeness, we also consider 
generally anisotropic hoppings in the three directions and 
anisotropic magnetic fluxes.
In Section \ref{non-Abelian} we introduce generic $U(2)$ configurations, also anisotropic, 
and derive the corresponding lattice Hamiltonian. We also deal with the 
definition of the MBZ for non-Abelian configurations. We find, 
in particular, that spatially-independent non-Abelian gauge potentials 
do not affect the structure of the MBZ, 
which turns out to depend only on the Abelian potential. The Hamiltonian in the Hasegawa 
gauge is explicitly written. As expected, the $SU(2)$ non-Abelian potential modifies 
the energy spectrum, leading in general to a splitting of the 
Abelian bands. In Section \ref{general} we investigate more formally 
the non-Abelian nature and the translational invariance of a 
given gauge potential, using the general properties of the Wilson loop. 
In Section \ref{examplestot} we perform a discussion of the single particle spectrum in the 
presence of both Abelian and non-Abelian isotropic potentials, the latter one  
mimicking a spin-orbit coupling, relevant in various proposals and experimental settings. In particular, we focus on the Abelian magnetic fluxes $\Phi= \pi, 2 \pi/3$, 
when the strength of the non-Abelian gauge coupling is continuously varied. 
Finally, in Section \ref{stab} we analyze the effects of small flux 
perturbations on the single particle spectrum. 
A mapping of these perturbed models to generalized Aubry-Andr\'e models 
is also described.  
We conclude the paper with an outlook on possible 
future developments and applications of the present work.

\section{Commensurate Abelian fluxes}
\label{Abelian}

The analysis of the physics of a particle in a three-dimensional lattice subject to magnetic fluxes has been a recurrent problem in the literature for many decades. Many authors addressed this problem adopting different approaches and focusing on several properties of this system, see, for example \cite{affleck,Hasegawa,Zou,MLT,LandauStat2,Zee1991,koshino2001,koshino2003}. In this Section we describe a suitable formalism for the analysis of such systems; we complete and extend the analysis in \cite{Hasegawa,LandauStat2} to pose the basis for the study of the non-Abelian gauge potentials in the following sections.

 We consider, in particular, a tight-binding model on a 
cubic lattice with $N=L^3$ sites and lattice spacing $a$,
the particles on the lattice being subject to an Abelian uniform and static magnetic field. 
In order to have on each plaquette (with area ${a}^2$) of the lattice 
a magnetic flux $\Phi = {B} {a}^2$, isotropic in the three directions, 
we consider a magnetic field $\vec{B} =  \Phi \, (1,1,1)$. 
In the next Subsection 
\ref{anisot} we deal with the case of anisotropic fluxes. 
The presence of such fluxes is connected to a phase $e^{i \Phi}$ for the 
hopping around a single plaquette. 
In presence of many species we can extend the subsequent treatment 
and results, given the fact that 
the Abelian gauge potential does not mix the different species. 

We write the commensurate magnetic flux as  
\beq
\Phi = 2 \pi \frac{m}{n} \, ,
\label{m_n}
\eeq 
with $m$ and $n$ red co-prime integers. The commensurability, reflected 
in the condition \eqref{m_n}, allows 
the analytical solution of the single particle spectrum (when 
the periodic boundary conditions are imposed) 
under the condition that the lattice encloses overall an integer number 
of fluxes in each direction, i.e. $L$ should be an integer multiple 
of $n$.  
In contrast, in the incommensurate case the spectrum can be reliably studied 
by rather heavy numerical computations on the real space 
tight-binding matrix \cite{lin96}. 

The magnetic field $\vec{B}$ can be put in connection with the 
gauge potentials $A_{\mu}$, with $\mu=0,1,2,3$. Choosing the 
Weyl gauge $A_0(\vec{x}) =0$, and following the usual 
formulation of a lattice theory in the presence of gauge potentials or fields 
(see for instance Ref.~\cite{dualat}), 
the real-space tight-binding Hamiltonian reads:
\begin{equation}
H = -  \sum_{\vec{r} \, , \, \hat{j}} \,  t_{\hat{j}} \, c^{\dagger}_{\vec{r} + \hat{j}} \,
e^{i \phi_{\vec{r} + \hat{j}, \vec{r}}}  \,
 c_{\vec{r}} \, + \ \mathrm{H.c.} \, ,
\label{peierls1}
\end{equation}
where the $t_{\hat{j}}$'s are the hopping amplitudes along the elementary 
displacements of the lattice, $\hat{j}=\hat{x}, \hat{y}, \hat{z}$. 
Periodic boundary conditions are assumed. This Hamiltonian constitutes a 3D extension of the Hofstadter model and its phases 
$\phi_{\vec{r} + \hat{j}, \vec{r}}$ are given by 
\begin{equation}
\phi_{\vec{r} + \hat{j} , \vec{r}} = \int_{\vec{r}}^{\vec{r} + \hat{j}} \vec{A}_{\text{AB} } (\vec{x}) \cdot \mathrm{d} \vec{x} \, ,
\label{peierls2}
\end{equation}
where $\vec{r}=(x,y,z)$ denotes the position of lattice sites. The 
subscript in $\vec{A}_{\text AB}$ indicates that we are considering an Abelian 
gauge potential, and from Section \ref{non-Abelian} onward a non-Abelian gauge 
potential will be added to it. 
Here and in the following, we fix the lattice spacing $a=1$ 
for the sake of simplicity, even though when useful we will restore it. 
As anticipated above, the action of the magnetic field is to make 
a particle on the lattice acquire a phase at every hopping process, 
the sum of these phases along a closed loop amounting precisely to the 
magnetic flux threading the surface bounded by the loop, in agreement with 
Stokes theorem. 

To study the effect of magnetic fluxes it is useful to 
recall the interplay between translational and gauge invariance 
in a system with a uniform magnetic field. We begin 
assuming a Hamiltonian in continuous space. Under this assumption, 
since the gauge potential is not constant like the related magnetic field, 
translational invariance implies that a translation of the coordinates 
by a vector $\vec{w}$ transforms the Hamiltonian as 
\begin{equation} 
\mathcal{T}^{\dagger}_{\vec{w}} (\vec{r}) \,  H(\vec{r} + \vec{w}) \, \mathcal{T}_{\vec{w}} (\vec{r}) =  H(\vec{r})  \, , 
\label{invab}
\end{equation}
with $\mathcal{T}_{\vec{w}} (\vec{r}) \in U(1)$ being a suitably 
chosen local gauge transformation which depends also on $\vec{w}$. 
This transformation acts on the Hamiltonian $H(\vec{r})$, linking the 
gauge potential at the point $\vec{r}$, $\vec{A}_{\text AB}(\vec{r})$, 
with the one at the translated point 
$\vec{r}+ \vec{w}$, $\vec{A}_{\text AB}(\vec{r}+ \vec{w})$:
\beq
\vec{A}_{\text AB}(\vec{r}) \to \vec{A}_{\text AB}(\vec{r}+ \vec{\omega}) = \vec{A}_{\text AB}(\vec{r}) - \, \vec{\nabla} \, \theta_{\vec{\omega}} (\vec{r}) \, ,
\label{gauge}
\eeq
with 
$\theta_{\vec{\omega}} (\vec{r})$ a scalar function 
(see for instance Ref.~\cite{peskin}). 
To determine the phases $\theta_{\vec{w}} (\vec{r})$, 
we consider that the vector potential, in the case 
of a uniform magnetic field, is linear in the space coordinates. Therefore 
we can write it as a function of a $3\times 3$ matrix $Q$: 
$$A_{{\text AB};j}(\vec{r})= Q_{ji} r_i \, ,$$
where $A_{{\text AB};j}$ is the $j$-th component of $\vec{A}_{\text AB}$ 
($j=x,y,z$).   
Under this assumption, the translation by $\vec{w}$ 
maps the vector potential into:
\begin{equation}
A_{{\text AB};j}(\vec{r})= Q_{ji} r_i \rightarrow A_{{\text AB};j}(\vec{r} + \vec{w}) = Q_{ji} r_i + Q_{ji} w_i\,.
\end{equation}
Therefore, to erase the contribution $Q_{ji} w_i$, based on Eq.~\eqref{gauge}, 
we must impose $\theta_{\vec{w}} (\vec{r})= r_j Q_{ji} w_i$. 
In this way, following the textbook approach \cite{LandauStat2}, 
we can define a magnetic translation operator $T_{\vec{w}}$ as the 
composition of the space translation $\vec{\omega}$ 
with the gauge transformation $\mathcal{T}_{\vec{w}} (\vec{r})$, 
characterized by $\theta_{\vec{w}}(\vec{r})$:
\begin{equation} \label{magtrasl}
T_{\vec{w}} \, \psi(\vec{r}) =  \mathcal{T}_{\vec{w}} (\vec{r})   \, \psi(\vec{r} + \vec{w}) = e^{-i\theta_{\vec{w}} (\vec{r})} \, \psi(\vec{r} + \vec{w}) = \psi(\vec{r}).
\end{equation}
The gauge redefinition of the wavefunction in Eq.~\eqref{magtrasl} 
is an example of Berry phase.
 
Coming back to the tight-binding model for a 
translationally invariant system on a cubic lattice, 
the latter transformation translates into:
\beq
H = -  \sum_{\vec{r} \, , \, \hat{j}} \,  t_{\hat{j}} \, c^{\dagger}_{\vec{r} + \hat{j}} \,
e^{i \theta_{\hat{j}}(\vec{r}) }  \,
 c_{\vec{r}} \, + \ \mathrm{H.c.} \, .
 \label{hamp}
\eeq
The related magnetic translation operators, 
$T_{\hat{x}}, T_{\hat{y}}$, and $T_{\hat{z}}$, do not commute with each 
other in general. But, in the case of commensurate fluxes, 
it is possible to find multiples of the unit vectors such that:
\begin{equation} 
\label{magtrasl2}
\left[T_{a\hat{x}}, T_{b\hat{y}} \right] = \left[ T_{a\hat{x}}, T_{c\hat{z}} \right] = \left[T_{b\hat{y}}, T_{c\hat{z}}\right] =0\,.
\end{equation}
A minimal triplet of integers $(a,b,c)$ of this kind defines a 
magnetic unit cell of volume $V_{uc}=abc$, playing a fundamental 
role in the definition of the MBZ. Indeed, the MBZ is 
defined by the reciprocal vectors (in quasi-momentum space) 
of three translations on the real lattice fulfilling the conditions in 
Eq.~\eqref{magtrasl2}.

We point out that in a general gauge the phases 
$e^{i \theta_{\hat{j}}(\vec{r}) }$ can be different from the ones 
defined in Eqs.~\eqref{peierls1} and \eqref{peierls2}, 
but the two sets are related by a gauge transformation as in 
Eq.~\eqref{gauge}. Moreover, Eq.~\eqref{peierls2} does not require 
translational invariance in general. Clearly, all 
the gauge-invariant quantities for the Hamiltonians in Eqs.~\eqref{peierls1} 
and \eqref{hamp} coincide, including the energies and the products
of the phases around a chosen closed path (the Wilson loop). Finally, the two sets of phases 
exactly coincide in the specific gauge 
$\vec{A} (\vec{r}) = \frac{1}{2} \, \vec{B} \times \vec{r}$ \cite{LandauStat2}.
The concept of  MBZ, just relying on translational invariance, 
can be defined for every gauge choice, starting from the phases 
defined as in Eqs.~\eqref{peierls1} and \eqref{peierls2}.

Due to the presence of the magnetic phases in Eq.~\eqref{peierls2}, 
the sites of the lattice, which are equivalent for $\vec{B} = 0$, are no longer 
equivalent. The lattice is then divided 
in a certain number of sublattices, and this division is gauge-dependent.  
This freedom may be exploited to individuate a gauge (or a set of gauges) 
giving rise to the smallest number of sublattices for the considered 
commensurate magnetic flux $\Phi = 2 \pi \frac{m}{n}$. It is clear 
that using the smallest number introduces a significant 
simplification in the computations. 

This set of gauges can be identified as follows. 
When considering a magnetic field $\vec{B}$ which is constant in space, 
then each component of the vector potential $A_i (\vec{r})$ must be at 
most linear in the space coordinates $\{x, y, z\}$. 
In this way, concerning the definition of the hopping phases, 
the point $\{x, y, z\}$ is equivalent to the point 
$\{x + \mathsf{a} n, y + \mathsf{b}n, z + \mathsf{c}n\}$, $\mathsf{a, b, c}$ 
being integers, meaning that the two points belong to the same sublattice. 
Moreover, either a hopping phase is constant along a direction $\hat{i}$, 
or at least $n$ values for it are required. 

Considering for a moment a two-dimensional square lattice, 
we conclude that the smallest number of its sublattices 
is $n$. This number is obtained, for instance, 
by setting to a constant ($0$ with no lack of generality) 
the magnetic phases along one direction and $\phi_l = 2 \pi \frac{m}{n} l$ 
($l=0, \dots ,n-1$) in the other one. A famous (and not unique) choice 
fulfilling these requirements is the Landau gauge 
$\vec{A}_{\mathrm{L}} = \Phi (0, x)$ \cite{LandauQM}, 
with $l = x \, \mathrm{mod} (n)$.

Let us focus now on a three-dimensional (3D) cubic lattice. 
In this case 
a single direction where the hopping phases are nonzero 
is clearly not sufficient, 
since the plaquette
orthogonal to this direction would have $\Phi=0$. Then, the best one 
can do is to keep $\phi_l = 2 \pi \frac{m}{n} l$ along only a direction and set 
$\tilde{\phi}_l = f_l (\phi_l)$ along another (with $f_l$ functions 
to be defined). In this way the number of sublattices still remains $n$.
Finally, the linear dependence of $A_i (\vec{r})$ on $\{x, y, z\}$ 
and the requirement of constant flux on all the plaquettes 
implies $\tilde{\phi}_l = \phi_{(n-l)} \, \mathrm{mod}(2 \pi)$. We conclude 
that the minimal number of sublattices is again $n$.

A gauge fulfilling the previous requirements and giving $n$ as 
dimension of the minimal Hamiltonian blocks is
\begin{equation} 
\vec{A}_{\text{AB} }  (\vec{r})= \frac{2\pi m}{n}   
 \, (0, x-y, y-x) \, 
\label{potgenab}
\end{equation}
(${x,y,z}$ can of course be permuted). 
The gauge \eqref{potgenab}, introduced by Hasegawa \cite{Hasegawa}, can be 
seen as a three-dimensional 
extension of the Landau gauge in two dimensions, and it reduces 
(up to a gauge redefinition) to the Landau gauge itself 
for $t_{\hat{z}} \to 0$. In particular, this gauge is known to simplify the three-dimensional model by reducing it to an effective one-dimensional problem in momentum space \cite{Zee1991}; this can be understood by observing that the explicit dependence on the position is a function of $x-y$ only, therefore there are two directions along which the momentum is conserved.

\begin{figure}[h!] \centering
\includegraphics[width=0.7\textwidth]{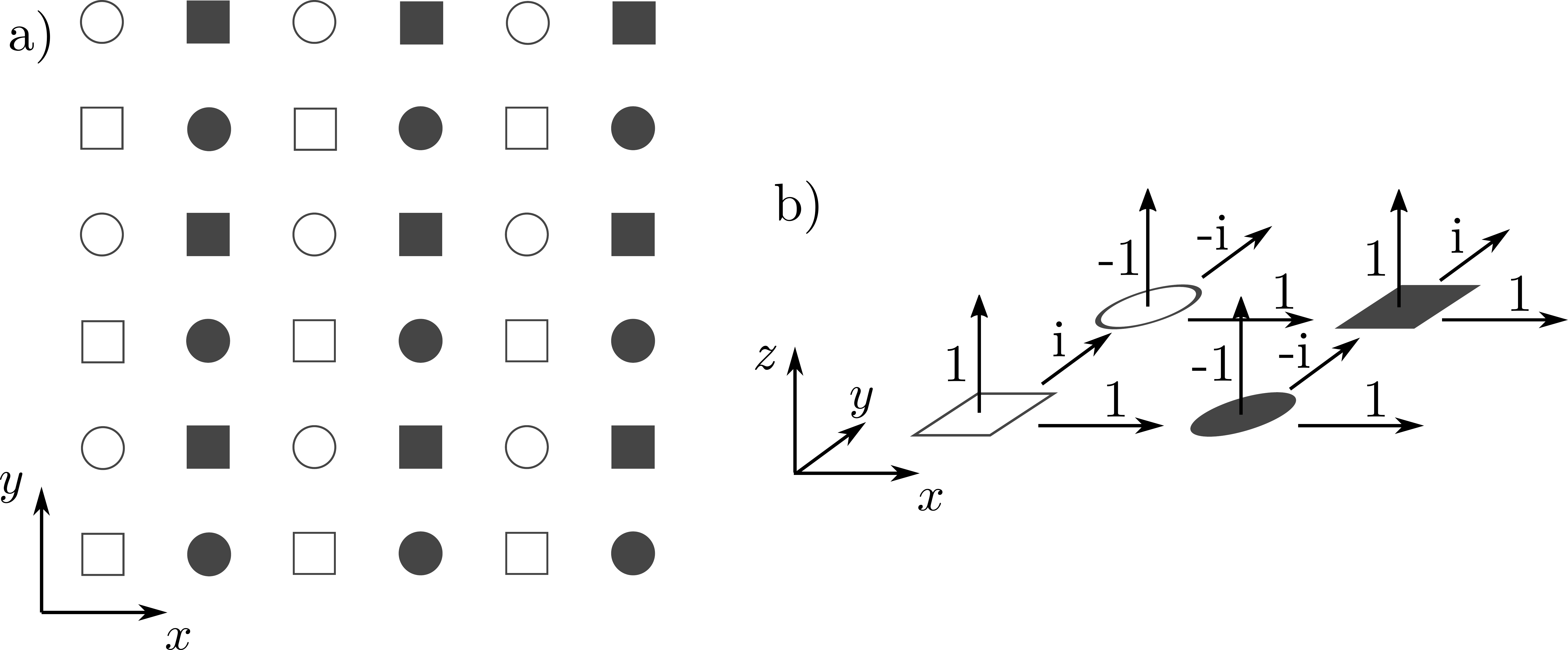}
\caption{a) Division of the lattice in sub-sublattices 
for the case $\frac{m}{n}  = \frac{1}{2}$ in the plane $(\hat{x}, \hat{y})$.  
The $n=2$ different sublattices characterized by the sets of hopping phases in Eqs.~(\ref{uy},\ref{uz}) are denoted by circles and squares. 
For each sublattice  it is convenient to define $n=2$ sub-sublattices which are represented as filled or empty 
symbols, respectively. b) The tunneling phases for $\frac{m}{n}  = \frac{1}{2}$ are depicted.}   
\label{lattice}
\end{figure} 

Using the choice \eqref{potgenab},
the Hamiltonian \eqref{peierls1} is rewritten in the form
\begin{equation} 
 H =  - \sum_{\vec{r}} \left[ t_{\hat{x}} \, c^\dag_{\vec{r}+\hat{x}} \, c_{\vec{r}} + 
t_{\hat{y}} \, U_{\hat{y}} (x,y) \, 
c^\dag_{\vec{r} + \hat{y}}c_{\vec{r}} +  t_{\hat{z}} \, U_{\hat{z}} (x,y) \, 
c^\dag_{\vec{r} + \hat{z}}c_{\vec{r}}  \right] + \, \mathrm{h.c.} \, ,
\label{Hab}
\end{equation}
where $U_{\hat{j}} (x,y) = e^{i \phi_{\vec{r} + \hat{j}, \vec{r}}}$. One finds 
$U_{\hat{x}} = 1$ and the following expressions for $U_{\hat{y}}$, $U_{\hat{z}}$: 
\begin{equation}
U_{\hat{y}} (x,y) = \exp\left[ {i \, 2 \pi \, \Big( x - y - \frac{1}{2} \Big) \, \frac{m}{n} } \right] \, , \label{uy} 
\end{equation}
\begin{equation}
 U_{\hat{z}} (x,y) =  \exp\left[ { - i \, 2\pi \, \Big( x - y \Big) \,  \frac{m}{n} } \right]  \, . 
 \label{uz}
\end{equation}
Notice that in our discussion we may set $\frac{m}{n} < 1$, 
since the spectrum is invariant for $\frac{m}{n} \to M - \frac{m}{n}$, 
with $M$ integer. We also observe that the $z$ coordinate is not 
present in Eq.~\eqref{potgenab}, so that the eigenfunctions can be written 
as $\psi(x,y,z)=e^{ik_z z}  \, u (x,y)$, with a dimensional reduction 
similar to the one occurring in two dimensions and giving rise to the 
Harper equation \cite{hofstadter76}.

The Hamiltonian \eqref{Hab} satisfies Eq.~\eqref{invab} and 
is translationally invariant. We stress that the 
property in Eq.~\eqref{invab} is a physical property of the system which is reflected in all the gauge-invariant observables, as for example, in the Wilson loops evaluated on closed paths along the lattice. Furthermore, the Hamiltonian 
\eqref{Hab} is also periodic (with period $n$) along the 
$\hat{x}$ and $\hat{y}$ directions, so that the 
wavefunctions $u (x,y)$ have the same periodicity.

Therefore it is possible to build a magnetic unit cell, defined by the elementary translations leading from a site to equivalent ones in the three lattice directions, such that it is enlarged $n$ times along both these directions, thus including $n^2$ sites. 

Consequently, the magnetic Brillouin zone (MBZ) is defined in momentum space as 
\begin{equation} [-\frac{\pi}{n},\frac{\pi}{n}) \times 
[-\frac{\pi}{n},\frac{\pi}{n}) \times [-\pi, \pi]\,.
\label{MBZ} 
\end{equation}
It is clear, however, that 
other permutations of the factors $\frac{1}{n}$ between 
space directions amount to a gauge redefinition of $\vec{A}(\vec{r})$, 
leaving the energy spectrum unaltered. For periodic boundary condition and with the choice \eqref{MBZ} 
there are 
${N}/{n^2}$ allowed momenta $\vec{k}$. Setting 
$\vec{k}=(k_x,k_y,k_z)$ with $k_x, k_y \in [-\frac{\pi}{n},\frac{\pi}{n})$,  and 
$k_z \in [-\pi,\pi)$, the components of $\vec{k}$ 
can be written as 
$k_x=\frac{2 \pi }{L} \, {\cal N}_x$, 
$k_y=\frac{2 \pi }{L} \, {\cal N}_y$ and $k_z=
\frac{2 \pi }{L} \, {\cal N}_z$, with ${\cal N}_{x,y} = 0, \dots, \frac{L}{n}-1$ and
${\cal N}_z = 0, \dots, L-1$,  
(similar expressions can be written if $N=L_x L_y L_z$ with $L_j$ 
the number of sites in the $\hat{j}$-direction). 

To find the $N$ eigenvalues of the Hamiltonian ~\eqref{Hab} on the considered 
cubic lattice with $N$ sites, one may exploit the division in sublattices 
just discussed. In particular we follow the approach in \cite{LandauStat2} to show that in the previously define MBZ, each eigenstate is $n-$fold degenerate. This is consistent with having $n$ bands, one for each sublattice, which are $n-$fold degenerate and include $N/n^2$ momenta.

 Using the Hasegawa gauge \eqref{potgenab}, 
the cubic lattice can be divided in $n$ sublattices, which 
we label by $s=(x-y) \, \mathrm{mod} (n)$. One then gets 
\begin{equation}
H = -  \sum_{\hat{j}} t_{\hat{j}} \sum_{s} e^{i \phi_{s, \hat{j}}}  \sum_{\vec{r}_{s}}  \, c^{\dagger}_{\vec{r}_{s} + \hat{j}} \, c_{\vec{r}_{s}} \, + \mathrm{H.c.} \, ,
\label{peierls3}
\end{equation}
where $s$ labels the different sublattices and $\vec{r}_s$ labels 
the sites of the $s$-th sublattice.

As a consequence, 
we expect $n$ sets of $\frac{N}{n}$ inequivalent energy eigenstates 
\cite{LandauStat2}, forming in general $n$ subbands in the MBZ. Any sublattice, however, 
is further divided in $n$ sub-sublattices differing by a 
translation $\pm (\hat{x} + \hat{y})$, which leaves invariant the potential 
\eqref{potgenab}. Therefore, any set of eigenstates is 
again partitioned in $n$ equivalent and degenerate sub-sets, 
and each one of these has $\frac{N}{n^2}$ elements. 
These elements are labelled by the $\frac{N}{n^2}$ MBZ momenta 
according the previous description, with the 
second partition leading to an $n$-fold degeneracy of each subband. 

In other words, with $\vec{k}$ belonging to the MBZ \eqref{MBZ} one has 
$N/n^2$ possible values of $\vec{k}$, while the tight-binding Hamiltonian 
\eqref{peierls1} has $N$ eigenvalues. For each of the $N/n^2$ 
values of $\vec{k}$ one has to diagonalize a $n \times n$ matrix, obtaining 
$N/n$ eigenvalues. Since each of them has a degeneracy $n$, we get the desired 
$N$ values for the eigenvalues of the  Hamiltonian 
\eqref{peierls1}. In Fig.~\ref{lattice} we pictorially 
represent the division in sublattices and sub-sublattices 
for the case $\frac{m}{n}  = \frac{1}{2}$.

We define 
$$c_{\vec{r}_s} = \frac{1}{\sqrt{N/n^2}} \, \sum_{\vec{k}} \, c_{s}(\vec{k}) \, e^{i \vec{k} \cdot \vec{r}_{s}}$$
with $\vec{r}_s$ indexing the $\frac{N}{n^2}$ 
sites of the sub-sublattice of the sublattice $s$. 
The Hamiltonian \eqref{peierls3} is then written as 
\begin{equation}
H = -  \sum_{\vec{k}} \sum_{\hat{j}} t_{\hat{j}} \sum_{s} e^{i A_{s, \hat{j}}} \,  e^{-i \vec{k} \cdot \hat{j}} \, c^{\dagger}_{s^{\prime}} (\vec{k}) \, c_{s} (\vec{k})\, + \mathrm{H.c.} \, .
\label{peierls4}
\end{equation}
In Eq.~\eqref{peierls4} we used $s^{\prime} \equiv s + \hat{j}$ to label the sublattice in which the particle moves into after the hopping; in particular $s$ and $s'$ coincide for the motion along $\hat{z}$ whereas they do not for the tunneling along $\hat{x}$ and $\hat{y}$, due to the potential \eqref{potgenab}.

The Hamiltonian \eqref{peierls4} 
can be written in matrix notation as  
\begin{equation}
H = - \sum_{\vec{k}} \sum_{\hat{j}} t_{\hat{j}} \sum_{s}  \, 
c^{\dagger}_{s^{\prime}}(\vec{k})  \Big(T_{\hat{j}}^{\mathrm{AB}} \Big)_{s^{\prime}, s} \,  e^{-i \vec{k} \cdot \hat{j}} \, c_{s}(\vec{k}) \, + \, \mathrm{H. c.} \, .
\label{hamab}
\end{equation}
The $T_{\hat{j}}^{\mathrm{AB}}$ are $n \times n$ matrices in the sublattice basis. 
Using the notation of Ref.~\cite{chapter}, they read:
\begin{align}
T_{\hat{x}}^{\mathrm{AB}} & = \begin{pmatrix}
        0 & 1 & 0 & 0 \\
        0 & 0 & \ddots & 0\\
        0 & \ldots & 0 & 1\\
        1 & 0 &  \ldots & 0
             \end{pmatrix}\,, \nonumber \\
  T_{\hat{y}}^{\mathrm{AB}} & = e^{- i \pi \frac{m}{n}} \, \begin{pmatrix}
        0 & \ldots & 0 & \varphi_0\\
       \varphi_{1}  & 0 & \ldots & 0\\
        0 & \ddots & 0 & 0\\
         0 & 0 &  \varphi_{n-1} & 0
      \end{pmatrix}\,,\nonumber \\
T_{\hat{z}}^{\mathrm{AB}} & =
                \begin{pmatrix}
        \varphi_0 & 0 & \ldots & 0\\
        0 & \varphi_{n-1} & 0 & 0\\
        0 & 0 & \ddots & 0\\
        0 & \ldots &  0 & \varphi_{1}
     \end{pmatrix}\,,
\label{matab}
\end{align}
where we introduced the notation 
$\varphi_{l} \equiv e^{i 2 \pi \frac{m}{n} l} \, , \, l = 0, \dots , n-1 $. 
We observe that the $n^{\rm th}$ power 
of the matrices $T_{\hat{x}, \hat{y},\hat{z}}^{\mathrm{AB}}$ 
gives the identity matrix. It is important to observe that 
when $\frac{m}{n} \neq \frac{1}{2}$, the matrices 
$T_{\hat{x}, \hat{y}, \hat{z}}^{\mathrm{AB}}$ are not invariant 
by the conjugate operation, expressing the fact that there is breaking 
of time-reversal symmetry. Finally, we observe that the previous results 
also apply in the presence of $p$ species 
(labeled by the index $\alpha = 1, \dots , p$) 
subject to the Abelian gauge potential in Eq.~\eqref{potgenab}. The Hamiltonian 
can be again written as $H = \sum_{\vec{k}} H(\vec{k})$ with 
\begin{equation}
H({\vec{k}}) = - \sum_{\hat{j}} t_{\hat{j}} \sum_{s} \, 
c^{\dagger}_{s^{\prime} = s + \hat{j}, \alpha^{\prime}} (\vec{k}) \left( T_{\hat{j}}^{\mathrm{AB}} \otimes  {\bf 1}_{p \times p} \right)_{s^{\prime}, \alpha^{\prime}, s, \alpha} \,  e^{-i \vec{k} \cdot \vec{j}} \, c_{s , \alpha} (\vec{k}) \, + \, \mathrm{H. c.} \, .
\label{hamab2}
\end{equation}

In conclusion, the diagonalization of a $N \times N $ matrix 
is reduced for a single degree of freedom 
to the diagonalization of a $n \times n $ one. Moreover, 
the spectrum is predicted to divide in $n$ subbands generally having 
different energies. This fact 
provides a decomposition of the original Hamiltonian into matrices 
having the minimal dimension.
In the particular case $\Phi = \pi$, two sub-bands are obtained, 
touching at Weyl cones as discussed in 
Refs.~\cite{LMT,lepori2015,ketterle2015}, providing 
the direct three-dimensional generalization of the square lattice model 
with $\pi$-fluxes discussed.

\subsection{Anisotropic Abelian lattice fluxes}
\label{anisot}

In the previous analysis we considered generally different hoppings 
$t_{\hat{j}}$ in the different space directions, but the same flux 
for the three orientations of the plaquettes. The latter 
condition can be relaxed to the case of 
anisotropic Abelian lattice fluxes. For the general magnetic field 
\beq
\vec{B} = 2 \, \pi \, \Big(\frac{m_x}{n_x}, \frac{m_y}{n_y},  \frac{m_z}{n_z} \Big) \, ,
\label{basim}
\eeq
it is convenient to choose the gauge potential
\beq
\vec{A}_{\text{AB}} (\vec{r})= 2 \pi  
 \, \Bigg(\Big(\frac{m_y}{n_y}-\frac{m_x}{n_x} \Big) \, (z-x), \frac{m_z}{n_z} \, (x-y), \frac{m_x}{n_x} \, (y-x) \Bigg) \, .
\label{potgenab3}
\eeq
One has then for the magnetic phases
\beq
\phi_{\vec{r} + \hat{j} , \vec{r}} =  2 \pi \, \Bigg(\Big(\frac{m_y}{n_y}-\frac{m_x}{n_x} \Big) \, \Big(z-x-\frac{1}{2}\Big), \frac{m_z}{n_z} \, \Big(x-y-\frac{1}{2}\Big), \frac{m_x}{n_x}  \, (y-x) \Bigg) \, .
\eeq
The important comment we would like to stress is that 
the gauge \eqref{potgenab3} again guarantees that there is 
the minimum number of (gauge-dependent) sublattices. In particular, 
we obtain $n_s$ inequivalent sublattices, with  
\beq
n_s = \mathrm{l. c. m. }(n_x, n_y, n_z) \,.
\label{VUC}
\eeq
Also in the considered case of anisotropic Abelian gauge potentials, 
each sublattice can be further divided in equivalent sub-sublattices 
following the same procedure detailed in the previous Section (se also 
\cite{libro}). 
Each inequivalent sublattice divides in $n_d$ equivalent sub-sublattices, with 
\beq
n_d = \mathrm{min} \Big(\mathrm{l. c. m. }(n_x,n_y) \, , \,  \mathrm{l. c. m. }(n_x,n_z) \Big) \, .
\eeq
There are $\frac{N}{n_s \, n_d} $ quasi-momenta defining each subband,
with $k_x = \frac{2 \pi }{L } \, {\cal N}_x$, $k_y=\frac{2 \pi }{L} \, 
{\cal N}_y$, and $k_z=\frac{2 \pi }{L} \, {\cal N}_z$, with 
${\cal N}_{x} = 0, \dots, \frac{L}{n_s}-1$, ${\cal N}_{y} = 0, \dots,\frac{L}{n_d}-1 $, and 
${\cal N}_{z} = 0, \dots, L-1 $. The Hamiltonian \eqref{peierls1} 
is then finally rewritten in the above defined MBZ and in the basis of 
the $n_s$ sublattices as in Eq.~\eqref{hamab}, and three 
$n_s \times n_s$ matrices are derived similarly 
to what has beed done in the previous Section for the 
three matrices $T_{\hat{j}}^{\mathrm{AB}}$ in Eq.~\eqref{matab}.
Similar results can be obtained using a different optimal gauge choice through the mapping onto a one-dimensional model in momentum space \cite{Zee1991}.

\section{Translationally invariant non-Abelian gauge potentials}
\label{non-Abelian}

We generalize here the Abelian models described in the previous Section, 
considering {\text two} species of particles hopping on the cubic lattice. 
For convenience, we label the effective spin degrees of freedom as 
$\Up$ and $\Dn$. In current ultracold atoms experiments, the two species can be obtained, for instance, by populating selectively two different hyperfine levels of a certain atom \cite{LibroAnna}. In addition to the commensurate magnetic field $\vec{B}$ with plaquette flux $\Phi = 2 \pi \frac{m}{n} $, we impose that these particles are subject to the effect of a non-Abelian, time-independent, translationally invariant $SU(2)$ gauge potential. Such a potential plays the role of a generalized spin-orbit coupling and it links in non-trivial ways the dynamics of the atoms with their spin, mixing the spin species.
From the point of view of the gauge group of the system, the models analyzed 
possess a gauge symmetry $U(2) \equiv \frac{SU(2) \times U(1)}{Z_2}$, 
which includes also the Abelian $U(1)$ symmetry group related with the 
conservation of the total number of particles $N_{\rm tot} = N_\Up + N_\Dn$ 
\cite{georgi}.  

The generating algebra of $U(2)$ is $u(2) = su(2) \oplus u(1)$. In the basis $(\Up,\Dn)$ the corresponding matrix representation can be taken as $\{{\vec{\sigma}, {\bf 1}}\}$, $\vec{\sigma}$ being the Pauli matrices. We define a generic $U(2)$ vector potential which is composed by the isotropic configuration of Abelian fluxes that we considered in the previous Section, combined with a translationally invariant non-Abelian $SU(2)$ contribution:
\begin{equation}  
\vec{A} (\vec{r})= \vec{A}_{\text{AB}} (\vec{r}) + \vec{A}_{\text{NAB}} \, .
\label{sum_A}
\end{equation}
The Abelian gauge potential is written as in the previous Section in the 
Hasegawa gauge: 
\begin{equation}  
\vec{A}_{\text{AB}} (\vec{r})= 2 \pi \, \frac{m}{n} (0, x-y, y-x)  {\bf 1} \, ,
\label{sum_A_1}
\end{equation}  
where ${\bf 1}$ is the $2 \times 2$ identity matrix. The non-Abelian 
gauge potential we consider reads
\begin{equation}  
\vec{A}_{\text{NAB}} (\vec{r})= 
(\vec{f}_x \cdot \vec{\sigma}, \vec{f}_y \cdot \vec{\sigma}, \vec{f}_z \cdot \vec{\sigma}) \, ,
\label{potabgen}
\end{equation}
where the vectors $\vec{f}_x, \vec{f}_y, \vec{f}_z$ do not depend explicitly 
on position. An additional term $(d_x , d_y , d_z) \, {\bf 1}$ in Eq.~\eqref{potabgen} can be erased by a gauge transformation.

The resulting tight-binding Hamiltonian, generalizing Eq.~\eqref{peierls3}, 
reads
\begin{multline}  
H = - \sum_{\vec{r}, \alpha, \alpha'} \left[t_{\hat{x}} \, U_{x,\alpha \alpha'}(\vec{r}) \, c^\dag_{\vec{r}+\hat{x},\alpha'} \, c_{\vec{r},\alpha} + t_{\hat{y}} \, U_{y,\alpha \alpha'}(\vec{r}) \, 
c^\dag_{\vec{r} + \hat{y},\alpha'}c_{\vec{r},\alpha} + \right.\\
\left. + t_{\hat{z}} \,  
U_{z,\alpha \alpha'}(\vec{r}) \, 
c^\dag_{\vec{r} + \hat{z},\alpha'}c_{\vec{r},\alpha}  \right] + \, \mathrm{H.c.} \, ,
\label{hamgen}
\end{multline}
where $\alpha$ and $\alpha'$ label the spin indices. In the considered case 
$U_{\hat{x}} \, , \, U_{\hat{y}}\, , U_{\hat{z}}$ are now matrices. 
$U_{\hat{x}}$ does not depend on the position, unlike $U_{\hat{y}}\, , U_{\hat{z}}$: 
we find
\begin{align}
&U_{\hat{x}} = \exp\left( {i\int_{x,y,z}^{x+1,y,z} A_x} \, \mathrm{d}x \right) = \exp\left( {i \, \vec{f}_x \cdot \vec{\sigma}}\right) \,, \label{uxnab} \\
 &U_{\hat{y}} (x,y) = \exp\left( {i\int_{x,y,z}^{x,y+1, z} A_y} \, \mathrm{d}y  \right) =\exp \left[ {i \, 2\pi \, \Big( x - y - \frac{1}{2} \Big) \, \frac{m}{n}  + i \, \vec{f}_y \cdot \vec{\sigma}} \right] \, , \label{uynab} \\
 &U_{\hat{z}} (x,y) =  \exp\left( {i\int_{x,y,z}^{x,y, z+1}A_z} \, \mathrm{d}z \right) = \exp\left[ { - i \, 2\pi \, \Big( x - y \Big) \, \frac{m}{n}   + i \, \vec{f}_z \cdot \vec{\sigma}} \right]  \,  \label{uznab}
\end{align} 
(remember that $a \equiv 1$). 
The Hamiltonian in Eq.~\eqref{hamgen} with the gauge potential in Eq.~\eqref{potabgen} is again translationally invariant because it fulfills Eq.~\eqref{invab}.  One can easily see that, since the non-Abelian term is position-independent, the translational invariance of the model is defined exactly as in the Abelian case and it is a gauge-invariant feature.
In the same way, the potential in Eq.~\eqref{potabgen} is genuinely non-Abelian, since it is not gauge equivalent to any other Abelian gauge potential. These general properties of the potential $\vec{A}$ will be discussed in more detail in the next section.

The gauge transformations of the system in Eq.~\eqref{hamgen}, with gauge 
potential given by Eqs.~\eqref{sum_A}--\eqref{potabgen}, 
are defined by the following unitary operators $\mathcal{U}$:
\begin{equation}
c_{\vec{r},\alpha} \to \mathcal{U}(\vec{r})_{\alpha,\alpha'}c_{\vec{r},\alpha'} \,, \quad \vec{A}(\vec{r})_{\alpha,\beta} \to  \mathcal{U}(\vec{r})_{\alpha,\alpha'} \vec{A}(\vec{r})_{\alpha',\beta'} \mathcal{U}^\dag (\vec{r})_{\beta',\beta}+
i \, \mathcal{U}^\dag(\vec{r})_{\alpha,\gamma}\left(\vec{\nabla} \mathcal{U}(\vec{r})\right)_{\gamma,\beta} \, ,
\label{nabtransf}
\end{equation}
where we sum over repeated indices. The previous transformations imply that the tunneling operators undergo the following transformation:
\begin{equation}
U_{\hat{j}}(\vec{r}) \to \mathcal{U}(\vec{r}+\hat{j}) U_{\hat{j}}(\vec{r}) \mathcal{U}^\dag (\vec{r}) \, .
\end{equation} 
Any local gauge transformation $\mathcal{U}$ leaves the Hamiltonian in Eq.~\eqref{hamgen} invariant.

The wavefunctions of  the same Hamiltonian can be again written as 
\begin{equation}
 \psi(x,y,z)=e^{ik_z z}  \, u (x,y) \, ,
\end{equation}
where $u (x,y)$ is  a periodic function with period $n$. 
For the same reason, the cubic lattice still divides in $n$ sublattices. The momentum-space Hamiltonian now takes the form
\begin{equation}
H = \sum_{\vec{k}} H(\vec{k}) 
\label{hamnab_prec}
\end{equation}
with \begin{equation}
H(\vec{k}) = 
- \sum_{\hat{j}} t_{\hat{j}} \sum_{s, \alpha, \alpha^\prime} \, 
c^{\dagger}_{s^{\prime} \equiv s + \hat{j}, \alpha^{\prime}} (\vec{k}) \left( T_{\hat{j}}^{\mathrm{AB}} \otimes  e^{i \, \vec{f}_j \cdot \vec{\sigma}} \right)_{{s^\prime}, \alpha^{\prime}, s, \alpha} \,  e^{-i \vec{k} \cdot \vec{j}} \,  c_{s, \alpha}  (\vec{k})  + \, \mathrm{H. c.} \, ,
\label{hamnab}
\end{equation}
where we used the same notation as in Eq.~\eqref{hamab2}. 
For generic values of $\vec{f}_x, \vec{f}_y, \vec{f}_z$,  the Hamiltonian in Eq.~\eqref{hamnab} has the same symmetries of the one in Eq.~\eqref{hamab}. As a consequence, its spectrum divides in $2n$ generally non-degenerate subbands. 
We finally notice that non-Abelian gauge potentials with $SU(2)$ 
group symmetry are the most general ones fully implementable
on a cubic lattice, since $SU(2)$ has only three independent generators. 
Gauge potentials with larger symmetry groups require non-cubic, 
more involved three-dimensional lattices.

The counting of eigenvalues goes as follows: the tight-binding 
Hamiltonian \eqref{hamgen}, defined for $N$ sites and $S=2$ inner degrees of 
freedom, has $SN$ eigenvalues. The possible values of ${\vec k}$ in 
the MBZ \eqref{MBZ} (unaltered by the presence of the non-Abelian terms) 
are $N/n^2$. For each of them one has to diagonalize 
a $Sn \times Sn$ matrix, for a total of $SN/n$ eigenvalues. Each of such 
eigenvalues has degeneracy $n$, giving the desired $SN$ eigenvalues. The same 
goes on for a Hamiltonian defined for general $S$ degrees of freedom 
(or components).

Eqs.~\eqref{hamnab_prec} and \eqref{hamnab} are 
the main result of the paper, since they show   
that in the presence of a commensurate Abelian gauge potential 
with flux $\Phi=2\pi \frac{m}{n}$ (with $m$ and $n$ integers) 
and of a general translational 
invariant non-Abelian gauge potential acting on a particle with two inner 
degrees of freedom, the diagonalization of the Hamiltonian can be reduced 
to the diagonalization of a $2n \times 2n$ matrix in the MBZ, with the MBZ 
unaltered by the non-Abelian terms. It is clear that if the particle has 
$S$ degrees of freedom, then the matrix to be diagonalized is $Sn \times Sn$. 
We finally observe that the previous treatment can be used 
for $t_{\hat{z}} \to 0$ to study the two-dimensional limit of the 
Hamiltonian in Eq.~\eqref{hamnab} with general non-Abelian gauge potentials.

In the following we give a couple of physically relevant applications 
of Eq.~\eqref{hamnab} in Sections 
\ref{examplestot} and \ref{stab}, but before we present in Section \ref{general} 
a general, more formal discussion of the properties of a 
non-Abelian gauge potential.

\section{General properties of a non-Abelian gauge potential}
\label{general}

In this section we review the rigorous definition of non-Abelianity of the gauge potential and we adopt it to study the effects of translational invariance in these systems. In particular, we show that the Brillouin zone of the 3D Hofstadter model is not affected by the introduction of a potential of the form  $\vec{A}_{\rm NAB}$ in Eq. \eqref{potabgen}.

The non-Abelian nature of this $U(2)$ potential seems evident from the fact that $\vec{A}_{\mathrm{NAB}}$ presents, in general, non-commuting components in the three directions. This implies that also the tunneling operators $U_{\hat{j}}\left(\vec{r}\right)$ in the site $\vec{r}$ do not commute with each other. Neither $\vec{A}_{\mathrm{NAB}}$ nor the operators $U_{\hat{j}}$, though, are gauge-invariant objects, and, as a result, also the commutators $\left[{A}_{\rm{NAB},\hat{j}},{A}_{\rm{NAB},\hat{k}}\right]$ and $\left[U_{\hat{j}},U_{\hat{k}}\right]$ depend on the gauge choice. This means that, in principle, there are seemingly non-Abelian gauge configurations that can be mapped into a fully Abelian case with $\left[{A}_{\rm{NAB},\hat{j}},{A}_{\rm{NAB},\hat{k}}\right]=\left[U_{\hat{j}},U_{\hat{k}}\right]=0$ with a suitable position-dependent gauge transformation in $U(2)$. The underlying model would thus be Abelian despite the conditions $\left[{A}_{\rm{NAB},\hat{j}},{A}_{\rm{NAB},\hat{k}}\right]\neq 0$ and  $\left[U_{\hat{j}},U_{\hat{k}}\right] \neq 0$ in the initial gauge choice. Therefore one needs a criterion to define a genuinely non-Abelian potential which cannot be mapped into an Abelian model with any gauge transformation.

For this purpose, it is useful to consider first the 
Wilson operator around a closed and oriented path $\mathcal{C}$:
\begin{equation} \label{Wline}
W(\mathcal{C},\vec{r}) = \mathbb{P} \, e^{i\oint_{\mathcal{C},\vec{r}} \vec{A}(\vec{r^{\prime}})\cdot d\vec{r^{\prime}}} \, ,
\end{equation}
where $\mathbb{P}$ denotes path ordering, required since the gauge potential matrices calculated at different points do not commute in general in the non-Abelian case, and $\vec{r}$ is the position of the initial and final site of $\mathcal{C}$, arbitrarily chosen.
While for an Abelian gauge configuration the operator $W(\mathcal{C})$ is gauge-invariant, related by the Stokes theorem 
to the magnetic flux on $\mathcal{C}$, and it is independent on $\vec{r}$, this is not so for a non-Abelian gauge configurations. Indeed in the latter case $W(\mathcal{C},\vec{r})$ transforms under the gauge group as \cite{Wei2}
\beq
 W(\mathcal{C},\vec{r}) \to \mathcal{U}^\dag(\vec{r}) \, W(\mathcal{C},\vec{r}) \, \mathcal{U}(\vec{r}) \, , 
 \label{transfW}
 \eeq
 Despite this gauge dependence, though, Wilson loops provide a sufficient criterion to define the genuine non-Abelian nature of the potential \cite{goldmanrev}. We may refer to a gauge potential as {\text genuinely} non-Abelian if there exist at least two closed 
paths $\mathcal{C}_1$ and $\mathcal{C}_2$ on the lattice, 
originating from the same site $\vec{r}$, 
such that the operators $W(\mathcal{C}_1,\vec{r})$ and $W(\mathcal{C}_2,\vec{r})$ 
do not commute with each other:
\begin{equation} \label{crit}
\left[W(\mathcal{C}_1,\vec{r}),W(\mathcal{C}_2,\vec{r})\right]\neq 0\,.
\end{equation}
Indeed, if \eqref{crit} hold, then $W(\mathcal{C}_1,\vec{r})$ and $W(\mathcal{C}_2,\vec{r})$  
cannot be put both in a diagonal form by the same gauge transformation $\mathcal{U}(\vec{r})$; therefore there exists no gauge choice in which the gauge potential can be written in a purely Abelian form.  We stress however that the latter criterium, although satisfactory from a mathematical point of view, results practically useless operatively, since exploiting it for a general gauge configuration leads to an exponentially (with the lattice size) hard problem. Instead, a different criterium overpassing this limit is still missing in our knowledge.

In an explicitly translational invariant system, condition \eqref{crit} can be simply verified for a single site, by looking at the minimal Wilson operators that describe the transport of an atom around a single lattice plaquette $\Box$; we define them as $W_{\hat{j}}(\Box)$, $\hat{j}$ labelling the orientation of the plaquette. The described situation holds for the potential in Eq. \eqref{potabgen}.
It is not possible to simultaneously diagonalize the three the 
plaquette operators $W_{\hat{j}}(\Box)$  through a gauge transformation: even though 
one can always find a gauge in which one of then is diagonal (writable as a combination 
of the identity matrix and $\sigma_z$), In this case
at least two closed 
lattice paths $\mathcal{C}_1$ and $\mathcal{C}_2$ exist, 
originating from the same site $\vec{r}$, 
such that \eqref{crit} is fulfilled, showing 
the intrinsic non-Abelian character of the potential \eqref{potabgen}.

In the case of the potential \eqref{potabgen}, its translational invariance is obvious due to its independence on the space coordinates. In general, though, the definition of translation invariance in the non-Abelian case requires more care. Indeed the qualitative difference between the Abelian and non-Abelian case lies again in the gauge-dependent behavior of the Wilson loops in the second case, as outlined by Eq. \eqref{transfW}.

For the Abelian potentials, 
a sufficient condition for translation invariance is that 
$W_{\hat{j}}(\Box)$ shall not depend on the initial and final site $\vec{r}$ of $\Box$, 
but only on the orientation of the plaquette, $\hat{j}$. In this case, 
by the Stokes theorem we have $W_{\hat{j}}(\Box)= e^{i \Phi_{\hat{j}} }$, 
where $\Phi_{\hat{j}}$ is a constant magnetic flux piercing each of the 
$\hat{j}$-oriented plaquettes.

For a non-Abelian gauge potential, instead, an explicit dependence of 
$W_{\hat{j}}(\Box)$ on the location $\vec{r}$ of $\Box$ does not necessarily imply the absence of translational 
invariance. A sufficient condition for translational invariance is indeed the existence of a gauge choice such that all the Wilson loops $W_{\hat{j}}(\Box)$ do not depend on $\vec{r}$. However, since the Wilson loops $W_{\hat{j}}(\Box)$ are not gauge invariant,
but transform as in Eq.~\eqref{transfW}, even if they are position independent in a specific gauge, they can acquire a non-trivial space dependence after a generic local gauge transformation $\mathcal{U}(\vec{r})$. 

In general, a Stokes theorem can be still formulated for $\mathrm{W}_{\hat{j}}(\Box)$ in the non-Abelian case 
\cite{stokes}, but the effective fluxes defined by it 
are not gauge invariant quantities.  
The physical origin of this fact 
is that, due to the non-linear nature of the non-Abelian gauge freedom 
(see for example \cite{Wei2}), the magnetic flux lines 
are themselves sources of flux. 

Because of the transformation in Eq.~\eqref{transfW}, 
we are then led to conclude that 
the Wilson loop itself cannot be used any longer to probe translational 
invariance in the non-Abelian case. 
When we consider $U(\mathcal{N})$ gauge potentials, though, each Wilson loop $W_{\hat{j}} (\Box)$ can be used to build $\mathcal{N}+2$ independent gauge invariant quantities: these are the trace, the determinant 
(equal to $1$ for the $SU(\mathcal{N})$ case, or 
equal to $e^{i  \mathcal{N}  \Phi}$ if an Abelian potential with 
flux $\Phi$ is also present), and the $\mathcal{N}$ minors of the $\mathcal{N}\times \mathcal{N}$ matrix describing $W_{\hat{j}} (\Box)$ \cite{lang}. 
A necessary condition for translational 
invariance along the axis $\hat{j}$ is the translational invariance of these $\mathcal{N}+2$ of these gauge-independent quantities for every plaquette 
$\Box$ oriented along $\hat{j}$.


The above condition is also sufficient: if a cubic lattice fulfills the previous condition, a space dependent gauge transformation 
can always be constructed such that all the Wilson loops $W_{\hat{j}}(\Box)$ become constant 
on every plaquette, thus making translational invariance explicit. 


For the gauge potential \eqref{sum_A}--\eqref{potabgen}, 
one has by an explicit computation $W_{\hat{x}}(\Box)=e^{i\Phi} 
e^{i \vec{f}_y \cdot \vec{\sigma}} e^{i \vec{f}_z \cdot \vec{\sigma}} 
e^{-i \vec{f}_y \cdot \vec{\sigma}}e^{-i \vec{f}_z \cdot \vec{\sigma}}$, 
$W_{\hat{y}}(\Box)=e^{i\Phi} 
e^{i \vec{f}_x \cdot \vec{\sigma}} e^{i \vec{f}_z \cdot \vec{\sigma}} 
e^{-i \vec{f}_x \cdot \vec{\sigma}}e^{-i \vec{f}_z \cdot \vec{\sigma}}$ and 
$W_{\hat{z}}(\Box)=e^{i\Phi} 
e^{i \vec{f}_x \cdot \vec{\sigma}} e^{i \vec{f}_y \cdot \vec{\sigma}} 
e^{-i \vec{f}_x \cdot \vec{\sigma}}e^{-i \vec{f}_y \cdot \vec{\sigma}}$. Therefore, 
we see that $W_{\hat{j}}(\Box)$ does not depend on the position in 
this particular case, and the condition that the 
$2+2$ invariants for $W_{\hat{j}} (\Box)$ are the same for every plaquette 
$\Box$ oriented along $\hat{j}$ is verified, 
showing the translational invariance.\\

Finally, we observe that the definition of the magnetic 
unit cell presented in Section \ref{Abelian} can instead be
directly extended to the case of the non-Abelian gauge symmetries. 

Analogously to \eqref{magtrasl}, we can define a magnetic translation operator $T_{\vec{w}}$ which is composed by the canonical translation operator $e^{i\vec{p} \cdot \vec{w}}$, where $\vec{p}$ is the momentum operator, and a gauge transformation $\mathcal{T}_{\vec{w}} \in G = G_{\rm AB} \times G_{\rm NAB}$ where we distinguish the Abelian and non-Abelian part of the gauge group (in the case under consideration $G=U(2)$). A gauge potential is translational invariant when it is  possible to define magnetic translation operators such that:
\begin{equation}
T_{\vec{w}} H T^\dag_{\vec{w}} = H\,, \quad {\rm with} \quad
T_{\vec{w}} \psi(\vec{r}) = \mathcal{T}_{\vec{w}} \psi(\vec{r} + \vec{w})
\label{magntrasnab}
\end{equation}
and $\mathcal{T} = \mathcal{T}_{\rm AB} \otimes \mathcal{T}_{\rm NAB}$.

Eq. \eqref{magntrasnab} implies that a position independent non-Abelian 
gauge potential $\vec{A}_{NAB}$ as \eqref{potabgen} does not influence the definition of the magnetic unit cell and Brillouin zone, that remains given by Eq.~\eqref{MBZ} 
as for the purely Abelian case in Section \ref{Abelian}. 
This peculiar result can be understood by observing that the non-Abelian contribution to the transformation $\mathcal{T}$ is not required: $\mathcal{T}_{\vec{w}} = \mathcal{T}_{\vec{w}}^{\mathrm{(AB)}} \otimes {\bf 1}$, therefore only the Abelian term must be considered to define the commuting translation operators $T$ in \eqref{magtrasl2}, and consequently the magnetic unit cell.

An explicit dependence of the MBZ on the non-Abelian gauge potential  could 
occur instead if the non-Abelian potential (assumed here not having other Abelian contributions) had a suitable spatial
dependence, for instance of the form 
\begin{equation}
 \vec{A}_{\mathrm{NAB}} = 2  \pi {\frac{v}{w}}  \left(x \, \hat{f}_x \cdot \vec{\sigma}, \; y \, \hat{f}_y \cdot \vec{\sigma},\; z \, \hat{f}_z \cdot \vec{\sigma} \right) \, ,
 \label{anabper}
 \eeq 
with $\hat{f}_i$ unitary vectors and $v,w$ integers. In this case the tunneling operators $U_{\hat{j}}$ are periodic with a period $w$ along the three directions, thus the Hamiltonian is explicitly invariant for translations of length $w$, therefore the magnetic translation operators $T_{w\hat{j}}$ become trivial.

\section{Isotropic non-Abelian gauge configurations} 
\label{examplestot}

In this Section we discuss specific examples of 
gauge configurations of the form Eq.~\eqref{potabgen}. 
To be specific, we consider a non-Abelian term of the form 
$(f_i)_j = q \delta_{ij}$ (with $i,j=x,y,z$) so that 
\begin{equation}  
\vec{A} = \vec{A}_{\text{AB}} + \vec{A}_{\text{NAB}} = 2 \pi \, \frac{m}{n} \, 
(0, x-y, y-x)  \, {\bf 1} + q \, (\sigma_x , \sigma_y, \sigma_z) \, ,
\label{potgen2}
\end{equation}
where the coupling $q$ takes continuous real values in the interval 
$[0, 2 \pi)$ (notice that there is invariance for $q \to q+2\pi$). 
The gauge potential \eqref{potgen2} 
describes the interplay between a constant Abelian magnetic field 
with magnetic flux $\Phi$ 
and a general $SU(2)$ spin-orbit coupling, extending the two-dimensional 
cases of the Rashba and Dresselhaus kind \cite{shenoy11,ryu2015,juze2017}.
An experimental proposal for its realization can 
be found in Ref.~\cite{anderson2012}. One can, of course, consider many 
other choices for $\vec{f}_x$, $\vec{f}_y$, $\vec{f}_z$. For instance, 
Ref.~\cite{lepori2016} uses $\vec{f}_x=q\,(1,0,0)$, 
$\vec{f}_y=q\,(0,1,0)$, and $\vec{f}_z=0$, leading to 
$\vec{A}_{\text{NAB}} = q \, (\sigma_x , \sigma_y, 0)$. The goal 
of the present Section is to illustrate specific examples of the rich 
structure deriving from the potential \eqref{potgen2}, in which a very 
symmetric -- isotropic and diagonal -- choice for the non-Abelian gauge 
potential is made.

It is easy to check that $|\mathrm{Tr} \, W_{\hat{j}}(\Box)| = 
2 \, (1- 2 \, \sin^2 q)$, such that the necessary condition 
$|\mathrm{Tr} \, W(\Box) |= 2$ found in Ref.~\cite{goldman09b} for 
the Abelian nature of a gauge configuration with constant $W(\Box)$ is 
fulfilled only in the two (gauge-equivalent) cases $q = 0, \pi$.

The tunneling matrices $U_{\hat{x}} \, , \, U_{\hat{y}} \, , U_{\hat{z}}$ in 
Eq.~\eqref{hamgen} are now given by:
\begin{align}
&U_{\hat{x}} = \exp\left( {i\int_{x,y,z}^{x+1,y,z} A_x \, \mathrm{d}x}\right) = \exp\left( {iq \, \sigma_x}\right) \,, \label{uxnab_bis} \\
 &U_{\hat{y}} (x,y) = \exp\left( {i\int_{x,y,z}^{x,y+1, z}A_y \, \mathrm{d}y}\right) =\exp\left[ {i \, 2\pi \, \Big( x - y - \frac{1}{2} \Big) \,  \frac{m}{n} + iq \, \sigma_y } \right] \, , \label{uynab_bis} \\
 &U_{\hat{z}} (x,y) =  \exp\left( {i\int_{x,y,z}^{x,y, z+1}A_z \, \mathrm{d}z}\right) = \exp\left[ { - i \, 2\pi \, \Big( x - y  \Big) \, \frac{m}{n} + iq \, \sigma_z } \right]  \, . \label{uznab_bis}
\end{align}
Similarly to the pure Abelian case in Eq.~\eqref{hamab}, 
the spectrum of the related Hamiltonian \eqref{hamnab} is invariant 
under $q \to - q$.

With the same notation of Eq.~\eqref{hamnab}, 
in the basis of the $n$ sublattices and in momentum space 
the Hamiltonian \eqref{hamgen} with the potential 
\eqref{potgen2} reads 
\begin{equation}
H = -\hspace{-0.5em} \sum_{\vec{k},\hat{j}, s, \alpha, \alpha^\prime} \hspace{-0.75em} t_{\hat{j}} \,  
c^{\dagger}_{s^{\prime} = s + \hat{j}, \alpha^\prime}  (\vec{k}) \left( T_{\hat{j}}^{\mathrm{AB}} \otimes  ( {\bf 1} \cos q  + i \sigma_{\hat{j}}  \sin q )\right)_{s^{\prime}, \alpha^{\prime}, s, \alpha} \hspace{-0.5em}  e^{-i \vec{k} \cdot \vec{j}}   c_{s, \alpha}  (\vec{k})  + \mathrm{\rm H. c.} \, ,
\label{hamnab2}
\end{equation}
where we used the relation 
$e^{i q \sigma_{\hat{j}}} = {\bf 1} \,\cos  q + i \,\sigma_{\hat{j}} \, \sin q$. 
We observe that in the two-dimensional limit $t_{\hat{z}} \to 0$, 
one recovers the family of topological insulators studied in Ref.~\cite{2Dsystem}. 

In the following we analyze the spectrum of the Hamiltonian for a few values 
of $n$ ($n=2,3,5$), considering for simplicity equal hopping amplitudes 
in all directions and denoting them by $t_{\hat{x}}=t_{\hat{y}}=t_{\hat{z}} 
\equiv t$.

\subsection{Abelian magnetic flux $\pi$}
\label{examples}

In this Subsection we analyze the case of Abelian magnetic flux $\pi$ 
(corresponding to $m=1$, $n=2$). 
The potential in Eq.~\eqref{potgen2} reads 
\begin{equation}
\vec{A} = \vec{A}_{\text{AB}} + \vec{A}_{\text{NAB}} = \pi \, (0, x-y, y-x) + q \,  (\sigma_x , \sigma_y, \sigma_z) \, .
\label{pot2}
\end{equation}
The unit cell of the system is composed of two subsets of sites (sublattice),
corresponding to even and odd $x-y$. Therefore, we can define an effective 
pseudospin-$1/2$ degree of freedom and a new set of Pauli matrices $\tau_i$ 
referring to it, with $i=x,y,x$. The tight-binding Hamiltonian 
\eqref{hamab} then reads
\begin{equation}
 H = \sum_{\vec{k},  s, s^{\prime}, \alpha, \alpha'} c_{s^{\prime},\alpha'}^\dag(\vec{k}) \, 
H_{s^{\prime} \alpha^{\prime}, s \alpha}(\vec{k}) \,  
\, c_{s, \alpha}(\vec{k}) \, , 
\label{sum_H}
\end{equation}
where $a, a^{\prime}$ label the eigenvalues of $\tau_z$. 
The matrix $H_{a^{\prime} \alpha^{\prime}, a \alpha}(\vec{k})$
is a $4 \times 4$ matrix involving direct products of Pauli matrices and it can 
be compactly written as
\begin{equation}
- \frac{{H}(\vec{k})}{2t}=  \cos q \cdot {\cal H}_0 + \sin q \cdot 
{\cal H}_1 \, ,
 \label{Hpiso}
\end{equation}
where we introduced the matrices
\begin{equation}
{\cal H}_0=\cos{k_x} \cdot \tau_x \otimes \sigma_0 + \cos{k_y} \cdot 
\tau_y\otimes\sigma_0 + \cos{k_z} \cdot \tau_z \otimes \sigma_0 = \sum_{\hat{j}} \cos{k_j} \left( 
\tau_j \otimes \sigma_0 \right) \, , 
\label{Hpiso_1}
\end{equation}
where to make uniform the notation we denoted by 
$\sigma_0$ the $2 \times 2$ identity matrix, and  
\begin{equation}
{\cal H}_1= \sin k_x \cdot \tau_x\otimes \sigma_x + \sin k_y \cdot 
\tau_y \otimes \sigma_y +  
\sin k_z \cdot \tau_z \otimes \sigma_z \, = \sum_{\hat{j}} \sin{k_j} \left( 
\tau_j \otimes \sigma_j \right) \, .
 \label{Hpiso_2}
\end{equation}
The explicit form for ${\cal H}_0$ and ${\cal H}_1$ are respectively 
\begin{equation}
{\cal H}_0 = \begin{pmatrix}\cos{k_z} & 0 & \cos{k_x}-i\cos{k_y} & 0 
\\ 0& \cos{k_z} & 0 & \cos{k_x}-i\cos{k_y} \\ \cos{k_x}+i\cos{k_y} 
& 0 & - \cos{k_z} & 0 \\ 0 & \cos{k_x}+i\cos{k_y} & 0 & \cos{k_z}\end{pmatrix} 
\label{H_0}
\end{equation}
and 
\begin{equation}
{\cal H}_1 = \begin{pmatrix}\sin{k_z} & 0 & 0& \sin{k_x}-\sin{k_y} 
\\ 0 & - \sin{k_z} & \sin{k_x}+\sin{k_y} & 0 \\ 0 & \sin{k_x}+\sin{k_y} 
& -\sin{k_z} & 0 \\ \sin{k_x}-\sin{k_y} & 0 & 0 & \sin{k_z}\end{pmatrix} \, . 
\label{H_1}
\end{equation}

The Hamiltonian in Eq.~\eqref{Hpiso} 
shows a discrete antiunitary particle-hole symmetry, defined by the matrix
\begin{equation}
 \mathcal{C}=\tau_y \otimes \sigma_y \, ,
 \label{phsym}
\end{equation}
so that
\begin{equation}
 H ( \vec{k} )  = - \, \mathcal{C} \, H^* ( -\vec{k} ) \, \mathcal{C}^{-1} \, .
 \label{phc}
\end{equation}
It is possible to show that the occurrence of the 
particle-hole symmetry is specific of 
the Abelian magnetic flux $\Phi = \pi$. 
Thus in general we can include this Hamiltonian in the class D 
(topologically trivial in three dimensions) of the classification of 
topological insulators and superconductors \cite{ludwig08}. 
This class is usually associated to Bogoliubov-de Gennes Hamiltonians 
describing superconductors, whereas in our case the particle-hole 
symmetry stems in a number conserving system from the $\pi$-fluxes 
in the lattice. Given our gauge choice, 
this particle-hole symmetry appears explicitly in the canonical level 
in the Hamiltonian Eq. \eqref{Hpiso}, whereas 
for other gauge choices one would need the addition of suitable gauge 
transformations to build a physical particle-hole symmetry. In a similar way, 
the system is also invariant under time-reversal symmetry, 
although its definition on the physical level requires a suitable 
space-dependent transformation \cite{lepori2016}.
 
Additional unitary $U(1)^3$ symmetries generated by the set $\{\tau_i \otimes \sigma_i\}$ appear if  $\frac{q}{\pi} = (2o+1)/2$, with $o$ an arbitrary integer 
(thus for $\cos q=0$).
In this case  $H(\vec{k})$ is real and has the further property $H(\vec{k})=-H(-\vec{k})$. Finally, if $\sin(q)=0$ is integer or half-integer 
the particle-hole like symmetry reduces to $\tau_y$, 
and the degree of freedom related with the two species of the hopping 
particles decouples from the Hamiltonian, leading to a further $SU(2)$ symmetry. 

Since the Hamiltonian in Eq.~\eqref{Hpiso} is a four by four matrix, 
its spectrum admits an involved closed-form analytic expression, whose explicit form 
we do not show. The spectrum divides into four sub-bands which, 
in general, overlap and touch, thus defining a metallic or semimetallic 
behavior of the system. The density of states as a function of $q$ is plotted 
in Fig.~\ref{lattice_DOS}, while the single particle 
ground state energy is reported in 
Fig.~\ref{GS}.

For $q=0$ the eigenvalues $E(\vec{k})$ 
of the Hamiltonian \eqref{Hpiso} are given (with degeneracy $2$) by 
\cite{Hasegawa}
\begin{equation}
\frac{E(\vec{k})}{2t} = \pm \sqrt{\cos^{2}{k_x}+\cos^{2}{k_y}+\cos^{2}{k_z}} \, ,
\end{equation}
so that the single particle ground state $E_0$ is 
$E_0=-2t \sqrt{3}$.

For $q=\pi/2$, instead, one has to diagonalize the matrix 
${\cal H}_1$ given in Eq.~\eqref{H_1}, obtaining the eigenvalues 
$$-\frac{E(\vec{k})}{2t}=\left\{ 
-\sin{k_x}\pm \left(\sin{k_y}+\sin{k_z}\right), 
\,\, \sin{k_x} \pm \left( \sin{k_y}-\sin{k_z} \right) \right\} \, , $$
with single particle ground state $E_0=-6t$. For general $q$, simple 
expressions are found when $|k_x|=|k_y|= 
|k_z|$. 
For $k_x=k_y=k_z\equiv {\cal K}$ the four eigenvalues 
of $H(\vec{k})$ are given by $-E(\vec{k}) / 2t=\pm \sqrt{3} \cos{q} 
\cos{{\cal K}}+\sin{q} \sin{{\cal K}}, \, -\sin{q} \sin{{\cal K}} 
\pm \sqrt{3 \cos^2{q} \cos^2{{\cal K}}+4 \sin^2{q} \sin^2{{\cal K}}}$, 
while for $-k_x=k_y=k_z\equiv \tilde{{\cal K}}$ they are 
$-E(\vec{k}) / 2t=\pm \sqrt{3} \cos{q} 
\cos{\tilde{\cal K}}+\sin{q} \sin{\tilde{\cal K}}, \, 
\sin{q} \sin{\tilde{\cal K}} 
\pm \sqrt{3 \cos^2{q} \cos^2{\tilde{\cal K}}+4 \sin^2{q} 
\sin^2{\tilde{\cal K}}}$. It is clear 
that for each quasimomentum $k_x=k_y=k_z={\cal K}$ there is a 
quasimomentum $-k_x=k_y=k_z=\tilde{\cal K}$ with the same four energy 
eigenvalues. We refer to the former set of quasimomenta 
$\vec{k}=({\cal K},{\cal K},{\cal K})$ with ${\cal K} \in [-\pi/2,\pi/2)$
and we choose $q$ by symmetry between 
$0$ and $\pi/2$. The states with minimum energy have 
$$
{\cal K}=-\frac{\pi}{2}
$$
as long as $q$ is larger than a value $q_c$ given by
$$
q_c=\frac{1}{2} \arccos{\left( \frac{1}{3} \right)}
$$
($q_c=0.6154797\cdots$) and smaller than $\pi/2$.  
As $q$ increases from $0$ to $q_c$, then ${\cal K}$ decreases 
from $0$ to $-\pi/2$. Namely for $q \in [0,q_c]$ 
the states with minimum energy in the set 
$\vec{k}=({\cal K},{\cal K},{\cal K})$ have 
$$
{\cal K}=-\arcsin{\sqrt{\frac{4 \cos^2{q} \sin^2{q}}{7-2 \cos{2q}+
7 \cos{4q}}}} \, ,
$$
and ${\cal K}=-\pi/2$ for $q \in [q_c,\pi/2]$.
The energy of these states is 
\begin{equation}
E_0({\cal K})=-2t \left( 
\frac{1}{2} \sqrt{6+7 \cos{2q}-\frac{1}{\cos{2q}}} +\sin{q} \, 
\sqrt{\frac{\cos^2{q} \sin^2{q}}{3 \cos^2{q} +4 \sin^2{q}-14 \cos^2{q} \sin^2{q}}} \right)
\label{sotto}
\end{equation}
for $q \in [0,q_c]$ and 
\begin{equation}
E_0({\cal K})=-6t \sin{q}  
\label{sopra}
\end{equation}
for $q \in [q_c,\pi/2]$. Eqs.~\eqref{sotto} and \eqref{sopra} are plotted 
in Fig.~\ref{GS}, from which one can see that the states 
$\vec{k}=({\cal K},{\cal K},{\cal K})$ (and their permutations as discussed) 
are indeed the single particle ground states of the Hamiltonian. 
One also has that for $q=q_c$ the ground state energy is 
$E_0=-2 t \sqrt{3}$, which is -- interestingly -- 
the same value as for $q=0$. The energy $E_0$ is non-monotonous with respect to $q$, showing a maximum at 
$q_{max} \approx 0.496$ [with $E_0(q_{max})\approx -3.281 t$] and then 
decreasing for $q$ larger than $q_c$.

\begin{figure}[h!]
\centering
\includegraphics[width=0.5\textwidth]{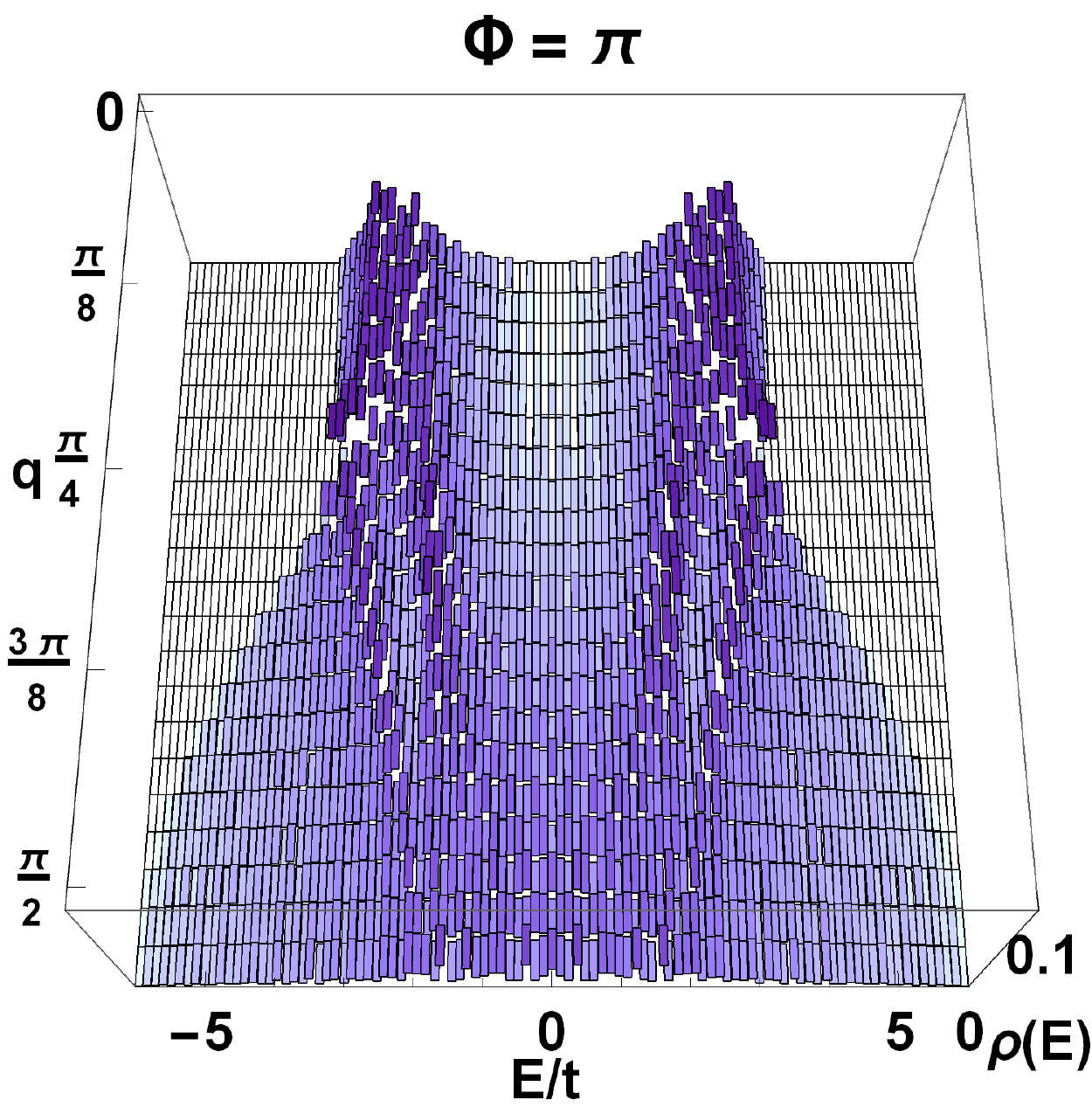}
\caption{Three-dimensional plot of the density of states $\rho(E)$  
(with the energy in units of $t$) 
as a function of the non-Abelian strength parameter $q$.}
\label{lattice_DOS}
\end{figure} 

\begin{figure}[h!]
\centering
\includegraphics[width=0.75\textwidth]{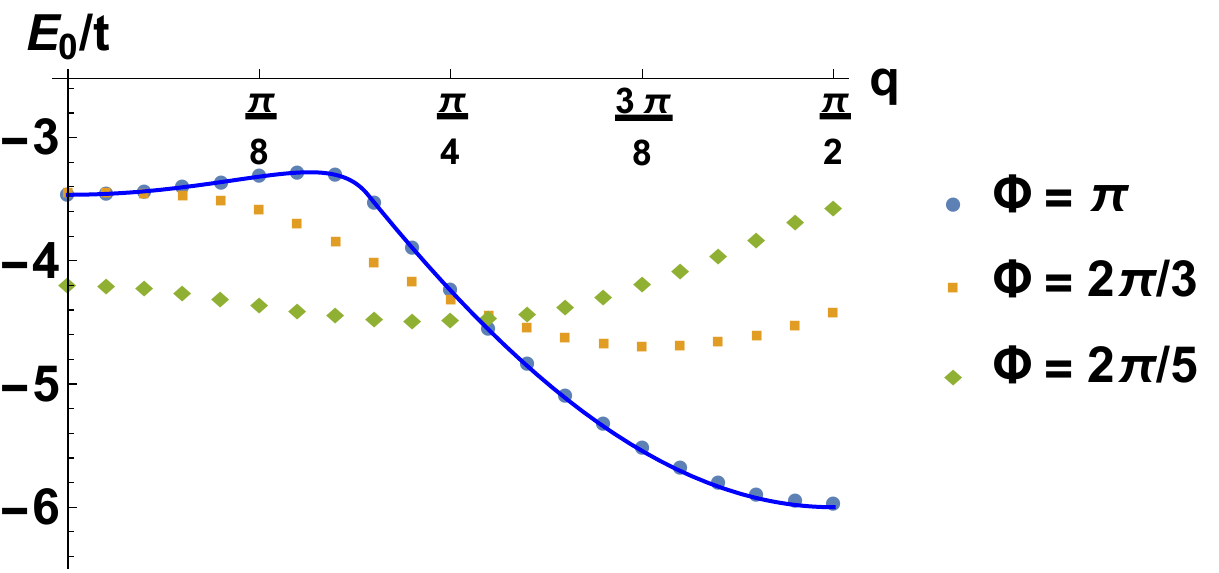}
\caption{Single particle ground state energy $E_0$ (in units of $t$) 
as a function of $q$ for $m=1$ and $n=2,3,5$. The solid line is the exact analytical solution for $n=2$. The dots are obtained by evaluating the spectrum in a discretization of the BZ into $25^3$ points.}
\label{GS}
\end{figure} 

For $m=1$ and $n=2$ we limited ourself to the case 
$\vec{A}_{\text{NAB}} = q \, (\sigma_x , \sigma_y, \sigma_z)$, 
but the previous results can be easily extended to non-Abelian anisotropic potentials 
of the form $\vec{A}_{\text{NAB}} = (q_x \,\sigma_x, q_y \, \sigma_y, q_z \, 
\sigma_z)$. An interesting case 
is obtained when $q_x=q_y \equiv q$ and $q_z =0$, which has been studied 
in relation to the formation of a double-Weyl semimetal phase 
\cite{lepori2016}. 

\subsection{Abelian magnetic flux $\frac{2 \pi}{3}$}

We focus in this Subsection on the cases with isotropic magnetic 
flux $\Phi = \frac{2\pi}{3}$, breaking explicitly 
the physical time-reversal symmetry. We consider the same non-Abelian 
gauge potential entering \eqref{potgen2}, so that the full potential reads
\begin{equation}
\vec{A} = \vec{A}_{\text{AB}} + \vec{A}_{\text{NAB}} = \frac{2 \pi}{3} \, 
(0, x-y, y-x) \otimes {\bf 1}+ q \,  (\sigma_x , \sigma_y, \sigma_z) \, .
\label{pot2a}
\end{equation}

With $\Phi = \frac{2 \pi}{3}$ the unit cell is composed of three sites, so it is useful to introduce a new pseudospin-$1$ 
degree of freedom with a diagonal operator labeled by $T_z^{\mathrm{AB}}$, 
whose diagonal entries characterize the 
$x$ coordinates of the lattices modulo $3$, as discussed in Section \ref{non-Abelian}. Correspondingly, 
the MBZ is defined by the quasi-momenta 
$k_x \in \left[-\pi,\pi\right)$ and $k_y,k_z \in \left[-\pi/3,\pi/3\right)$. 
By introducing the matrices
\begin{equation}
T_{\hat{x}}^{\mathrm{AB}} = \begin{pmatrix}0& 0 & 1 \\ 1& 0 & 0 \\ 0& 1 & 0 \end{pmatrix}\, , \quad   T_{\hat{y}}^{\mathrm{AB}} = \begin{pmatrix} 0 & 0 & e^{-i \pi/3}  \\ e^{i \pi/3} & 0 & 0  \\ 0 & -1 & 0 \end{pmatrix}  \, \quad  
T_{\hat{z}}^{\mathrm{AB}} = \begin{pmatrix} 1 & 0 & 0 \\ 0 & e^{i\frac{4\pi}{3}} & 0 \\ 0 & 0 & e^{i\frac{2 \pi}{3}} \end{pmatrix} \, ,
\end{equation}
the $6 \times 6$ Hamiltonian in Eq.~\eqref{hamab} reads in momentum space:  
\begin{equation} 
-\frac{H(\vec{k})}{t} = 
\cos q \cdot \sum_{\hat{j}} e^{- i k_j} T_j^{\mathrm{AB}} \otimes 
\sigma_0 +  i\sin q \cdot \sum_{\hat{j}} e^{- ik_j} T_j^{\mathrm{AB}} \otimes \sigma_j+ {\rm H. c.} \, ,
\label{ham3}
\end{equation} 
where $\hat{j}=\hat{x},\hat{y},\hat{z}$. To recast Eq.~\eqref{ham3} in a form 
closer to Eq.~\eqref{Hpiso}, we write the matrices $T_{\hat{j}}^{\mathrm{AB}}$ 
as $T_{\hat{j}}^{\mathrm{AB}}=R_{\hat{j}}^{\mathrm{AB}}+S_{\hat{j}}^{\mathrm{AB}}$, 
where $R_{\hat{j}}^{\mathrm{AB}}$ is Hermitian and $S_{\hat{j}}^{\mathrm{AB}}$ is 
anti-Hermitian so that $R_{\hat{j}}^{\mathrm{AB}}=(T_{\hat{j}}^{\mathrm{AB}}+T_{\hat{j}}^{\mathrm{AB} \dag})/2$ and $S_{\hat{j}}^{\mathrm{AB}}=(T_{\hat{j}}^{\mathrm{AB}}-T_{\hat{j}}^{\mathrm{AB} \dag})/2$. In this way we can write Eq.~\eqref{ham3} as 
\begin{equation} 
-\frac{H(\vec{k})}{2t} = \cos q \cdot {\cal H}_0 + \sin q \cdot 
{\cal H}_1 \, ,
\label{ham4}
\end{equation} 
where
\begin{equation}
{\cal H}_0=\sum_{\hat{j}} \left\{ \cos{k_j} \left( 
R_{\hat{j}}^{\mathrm{AB}} \otimes \sigma_0 \right) + \, \sin{k_j} \left( 
I_{\hat{j}}^{\mathrm{AB}} \otimes \sigma_0 \right) \, \right\}, 
\label{Hpiso_1_3}
\end{equation}
and   
\begin{equation}
{\cal H}_1= \sum_{\hat{j}} \left\{ \sin{k_j} \left( 
R_{\hat{j}}^{\mathrm{AB}} \otimes \sigma_j \right) - \, \cos{k_j} \left( 
I_{\hat{j}}^{\mathrm{AB}} \otimes \sigma_j \right) \, \right\} \, ,
 \label{Hpiso_2_3}
\end{equation}
where we introduced the Hermitian matrices $I_{\hat{j}}^{\mathrm{AB}} \equiv -i 
S_{\hat{j}}^{\mathrm{AB}}$. We expect the structure defined 
in Eqs.~\eqref{ham4}--\eqref{Hpiso_2_3} to be valid 
for general values of $m$ and $n$. Moreover, the comparison between Eqs.~\eqref{ham4}--\eqref{Hpiso_2_3} and Eqs.~\eqref{Hpiso}--\eqref{Hpiso_2} show the peculiarity of the $\pi$-flux case, where the 
matrices $T_{\hat{j}}^{\mathrm{AB}}$ are Hermitian, so that 
$I_{\hat{j}}^{\mathrm{AB}}=0$ [indeed the matrices 
$T_{\hat{j}}^{\mathrm{AB}}$ are just the Pauli matrices for $m=1$ and $n=2$].

An interesting point to be noticed is that for $\Phi=\pi$ and $q=0$, the single particle 
ground state is obtained for $k_x=k_y=k_z=0$, giving $E_0=-2\sqrt{3}t$. For 
flux $2 \pi/3$, choosing $q=0$ and setting $k_x=k_y=k_z=0$ one gets 
six energies, and the smallest is given by $-\frac{1}{2} 
\left( 1+\sqrt{33} \right) t\approx -3.37t$, which appears to be above 
the single particle ground state energy $E_0$ found by diagonalizing 
Eq.~\eqref{ham4} on all the MBZ without the restriction $| k_x| = | k_y| = | k_z|$. The position of the minimum in momentum space 
is a non-trivial problem and it is obviously related to the gauge choice. 
We observe indeed that by multiplying any of the $T_{\hat{j}}^{\mathrm{AB}}$ 
matrices by a phase, the spectrum is translated accordingly in momentum space, 
even though of course the values of the eigenenergies do not change. 
It would certainly be interesting to determine 
a choice of the gauge simplifying  -- or fixing, if such choice does exist -- 
the determination of the position of the minimum.

Qualitatively, 
for $q=0$ the spectrum of the Hamiltonian~\eqref{ham3} is composed 
of three doubly-degenerate bands, due to the absence of any spin term. 
These bands touch in points with a linear dispersion, possibly 
realizing a Weyl semimetal phase at fillings $2/3$ and $4/3$. 
This is reflected in the quadratic behavior of the density of states 
depicted in Fig.~\ref{fig:DOS} around the two minima, 
at zero density for $q=0$, which separate the three degenerate bands. 
For $q>0$ the spectrum divides into six bands which are still connected, 
in general, by band-touching points but, otherwise, they do not intersect. 
For small values of $q$, the six bands do not overlap in energy, 
thus determining, in general, metallic phases, separated by semimetal phases 
at commensurate filling $n/3$. Above a critical value, though, the bands 
start to overlap and the system is in a metallic phase for any filling. 
Finally, for $q$ approaching $\pi/2$, the density of states becomes 
again suppressed in small energy ranges separating the bands. 
Also in this case, these regions do not correspond to gaps between 
the bands but to semimetallic phases with band touching points between them. 
We observe that the energy bands display no particular symmetries, 
as expected from the Hamiltonian in Eq.~\eqref{ham3}. 

A similar picture emerges for larger $n$: e.g., for $m=1$ and $n=5$, 
corresponding to $\Phi = \frac{2 \pi}{5}$. For $q = 0$ 
five twofold-degenerate bands occur, that split in ten single bands 
at $q \neq 0$. The density of states $\rho (E)$ vanishes at isolated 
points along the lines $q = 0$  and $q = \frac{\pi}{2}$. At the corresponding 
energies, we observe a linear band touching between neighboring bands.

\begin{figure}[h!]
\centering
\includegraphics[width=0.5\textwidth]{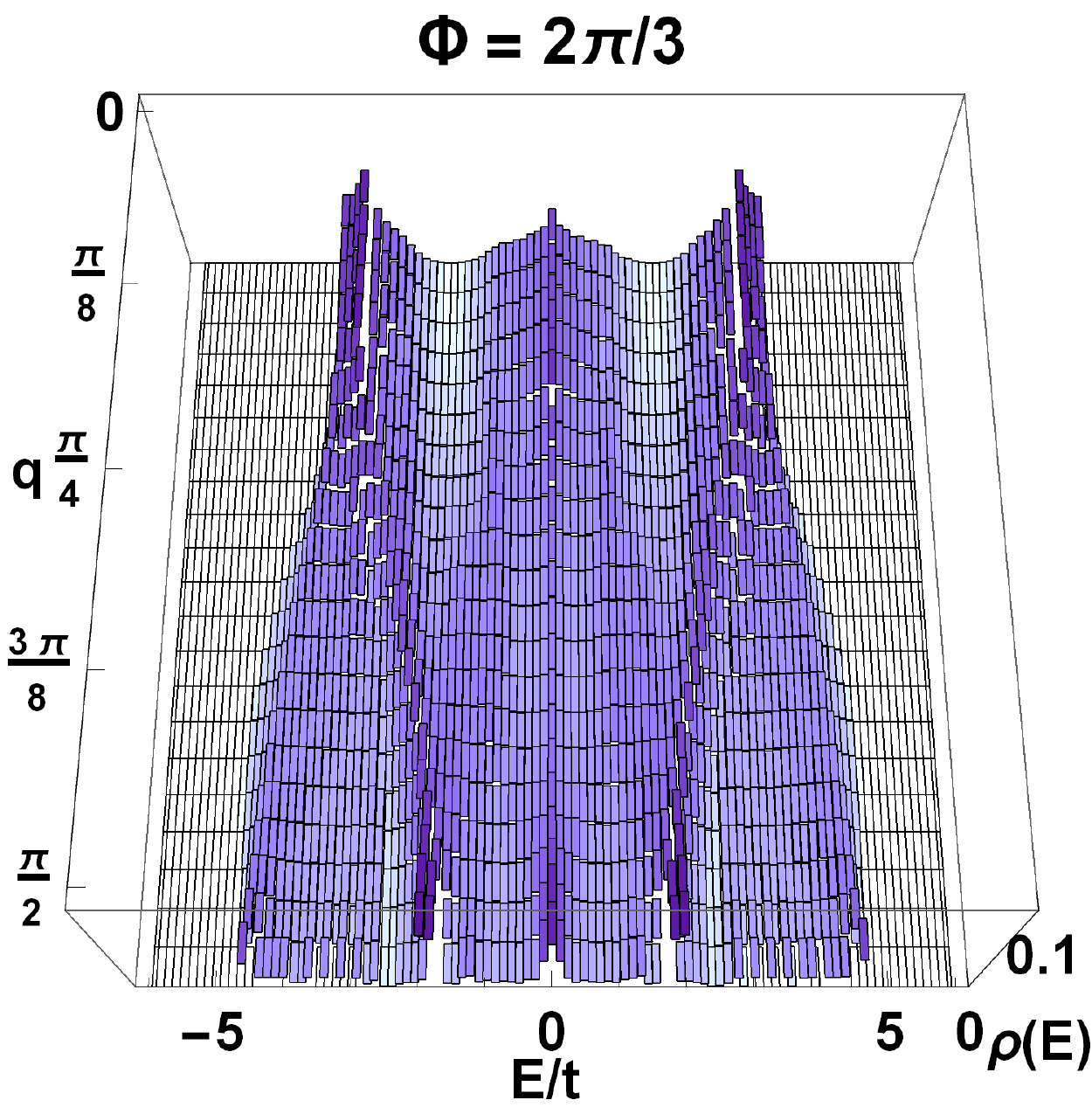}
\caption{Density of states of the Hamiltonian in  Eq. \eqref{ham3} as a function of the energy and the parameter $q$. For $q=0$ two regions with the DOS vanishing quadratically are particularly evident: they correspond to linearly dispersing band touching points.}
\label{fig:DOS}
\end{figure}

\section{The effect of flux perturbations}
\label{stab}

The definitions of the magnetic unit cell and Brillouin zone discussed 
in the previous Sections can be used to shed light on the behavior of systems 
with artificial gauge potentials in finite size systems, also when the Abelian 
fluxes are slightly perturbed around some key values, and, in particular, 
around a configuration where the Abelian potential defines $\pi$ fluxes for 
all the three orientations of the plaquettes of the cubic lattice.

In Refs.~\cite{LMT,ketterle2015} it was shown that the cubic lattice model 
with $\pi$ fluxes hosts a Weyl semimetal phase and in Ref.~\cite{lepori2015} 
we discussed the effect of small random perturbations around this value 
of the fluxes. Interestingly, the results for small system sizes are 
analogous to those obtained by the introduction of onsite disorder 
in solid-state realizations of Weyl semimetals \cite{pixley2016, Sbierski14, Sbierski16, Trescher16, Bera16, Roy16}. 
For small flux perturbations, the density of states of the system 
at low energy shows a deviation from the linear behavior typical of the Weyl 
semimetals, which is compatible with the introduction of rare 
localized states. 

To address a specific and physically relevant 
example in the formalism introduced in the present 
paper, we can consider, in the thermodynamic limit, the system 
with random fluxes of the form $\pi (1\pm 1/p)$, with $p$ integer and 
$1/p$ a small perturbation around $\pi$. The system then display 
a fractal spectrum (see for example Refs.~\cite{Zee1991,koshino2001,koshino2003}), 
and the Weyl physics disappears.
This leads to instabilities even for infinitesimal fluctuations of the 
Abelian flux, in a context that it is similar to its two-dimensional 
counterpart, described by the Hofstadter butterfly \cite{hofstadter76}.

The introduction of the parameters $p_i$ in the three directions, however, 
causes the appearance of a volume scale given by the size of the magnetic 
unit cell, which grows with the least common multiple of the $p_i$ parameters, 
as dictated by Eq.~\eqref{VUC}. Therefore, the thermodynamic behavior 
describes only systems larger than this unit cell. For smaller sizes 
we will show in the following that the system is analogous to a collection 
of disordered two-dimensional models, which explains the physics of 
finite sizes and small perturbations.

To describe these systems in more detail, we consider the case of a 
gauge potential displaying $\pi$ magnetic fluxes along the three 
directions and a non-Abelian component which is gauge-equivalent 
to the common two-dimensional spin-orbit couplings (Rashba or Dresselhaus). 
We emphasize that such a non-Abelian term is analogous to the one recently 
experimentally realized with $^{40} K$ gases in continuum space 
\cite{huang2016}, and a recent proposal paves the way for its realization in optical lattices \cite{demler2017}. The gauge potential reads
\begin{equation}  
\vec{A} = \vec{A}_{\text{AB}} + \vec{A}_{\text{NAB}} = \pi \, (0, x-y, y-x) \, {\bf 1} + (q_x\sigma_x , q_y\sigma_y, 0) \, ,
\label{potrashba}
\end{equation}
with
\begin{align}
&U_{\hat{x}} =  \exp\left( {iq_x \, \sigma_x}\right) \,, \label{uxnab2} \\
 &U_{\hat{y}} (x,y)  =\exp\left[ {i \, \pi \, \Big( x - y - \frac{1}{2} \Big)  + iq_y \, \sigma_y } \right] \, , \label{uynab2} \\
 &U_{\hat{z}} (x,y) =   \exp\left[ { - i \, \pi \, \Big( x - y  \Big) } \right]  \, . \label{uznab2}
\end{align}

In the limit $q_x,q_y \to 0$, the system describes a 
PT-invariant Weyl semimetal \cite{lepori2015}. In the symmetric case 
$q_x=q_y$, in Ref.~\cite{lepori2016} it was shown that it corresponds 
to a double-Weyl semimetal, namely a gapless system in which 
the central bands touch in points whose dispersion is quadratic along 
$\hat{x}$ and $\hat{y}$ and linear along $\hat{z}$. Such band touching 
points are topological objects characterized by a double monopole 
of the Berry curvature and they are protected by the $C_4$ rotational 
symmetry of the system \cite{bernevig12}. By introducing an 
anisotropy with $q_x \neq q_y$, the double-Weyl points 
split into pairs of Weyl cones with the same charge.

To understand the role of the Abelian flux perturbations 
in the cubic lattice model it is convenient to use the following gauge 
choice for the Abelian component of the vector potential:
\begin{equation} \label{genericfluxes}
\vec{A}_{AB}=\left( 0,  \gamma x - \pi y, \alpha y - \beta x\right)\,.
\end{equation} 
This Abelian component corresponds to a generalization of the Abelian 
contribution in Eq.~\eqref{potabgen} such that the magnetic fluxes are 
$(\alpha,\beta,\gamma)$, corresponding to a generic choice of the $U(1)$ 
fluxes in all the plaquettes of the cube. The potential in 
Eq.~\eqref{genericfluxes} does not depend on the $z$ coordinate due 
to this gauge choice. This allows us to consider the tight-binding 
Hamiltonian obtained from this perturbation of the fluxes and the 
Rashba-like spin-orbit coupling in Eq.~\eqref{potrashba} as a function of 
the real space coordinates $x,y$ and the momentum $k_z$, which may be 
thought of as a parameter labeling different two-dimensional systems in 
the $xy$ plane. 

Therefore, we can rewrite the Hamiltonian as:
\begin{multline} \label{harper2D}
H = - \sum_{k_z} \sum_{x,y} \left\{\left[t_{\hat{x}} U_{\hat{x}}(x,y) c^\dag_{x+\hat{x},y,k_z}  c_{x,y,k_z} + t_{\hat{y}} U_{\hat{y}}(x,y) c^\dag_{x,y+\hat{y},k_z}  c_{x,y,k_z}+{\rm H. c.}\right] +\right.\\
+ \left. 2t_{\hat{z}}\cos\left(k_z + \alpha y - \beta x \right) c^\dag_{x,y,k_z}  c_{x,y,k_z}\right\} \, ,
\end{multline}
where the unitary operators $U_{\hat{x}}$ and $U_{\hat{y}}$ include both the 
action of the Abelian flux $\gamma$ piercing the plaquettes of the 
two-dimensional $x-y$ system and the non-Abelian hopping operators along 
the $x,y$ directions:
\begin{equation}
U_{\hat{x}}=e^{i q_x \sigma_z} \,,\quad
U_{\hat{y}}=e^{i\left(\gamma x -\pi y -\frac{\pi}{2}\right) + iq_y \sigma_y}\,.
\end{equation}
These operators describe the kinetic contribution of 
the two-dimensional Hamiltonian. The last term in Eq.~\eqref{harper2D} 
can be considered as an onsite potential in the plane $x-y$ oscillating 
in space and depending on the values of $\alpha$ and $\beta$, 
with $k_z$ being just a phase for this periodic potential. 
This oscillating potential characterizes as well the so-called 
Aubry-Andr\'e model \cite{aubry}. When $\alpha$ and $\beta$ 
are incommensurate with the optical lattice spacing, the Hamiltonian 
\eqref{harper2D} describes a two-dimensional system of particles subject 
to $\pi$ fluxes and spin-orbit coupling and moving in a quasi-periodic system.
In this incommensurate regime, the previous potential 
can drive a transition from extended to localized states. 

In finite systems, commensurate potentials may show the same behavior 
as incommensurate ones, when their spatial period is large compared 
to the system size. We consider here what happens if $\alpha$ and $\beta$ 
are weakly perturbed around the original value $\pi$. We assume their 
value is of the kind $\alpha,\beta = \pi (1\pm 1/p_{\alpha/\beta})$ with odd 
integers $p_\alpha,p_\beta \gg 1$. 
The spatial period of the onsite potential in the plane becomes 
$p_\alpha$ in the $\hat{y}$ direction and  $p_\beta$ along $\hat{x}$. Therefore, 
if the system has a size $L_{x,y} \ll p_{\beta,\alpha}$, the system cannot 
be truly considered in the thermodynamic limit because 
its size is considerably smaller that the period of the onsite potential, 
and its phenomenology reproduces the one of an incommensurate potential. 
This effect has been experimentally investigated with Bose-Einstein 
condensates \cite{roati2008}, where it was shown that the introduction 
of a quasi-periodic potential on a finite system can indeed cause a 
crossover to a regime with localized states. 

Therefore, for small perturbations of the Abelian fluxes on a finite system, 
the Hamiltonian in Eq.~\eqref{harper2D} can be considered, 
for each value of $k_z$, as a two-dimensional Hofstadter model with flux 
$\gamma$ and spin orbit terms dictated by $U_{\hat{x}}$ and $U_{\hat{y}}$ 
\cite{2Dsystem} with the addition of an effective on-site disorder, 
characterized by the phase $k_z$. This qualitatively explains the 
appearance of a diffusive phase in the perturbed Abelian system 
\cite{lepori2015}, behaving analogously to a disordered Weyl semimetal and
characterized by the appearance of rare quasi-localized states 
\cite{pixley2016}. We observe that this mapping from a three-dimensional 
cubic lattice model with fluxes to a two-dimensional 
square lattice model with a quasi-periodic potential follows 
the usual mapping from the two-dimensional Harper model 
to the one-dimensional Aubry-Andr\'e model. In the following we study 
the stability of the energy spectrum in presence of gauge fluctuations.

\subsection{Stability of the energy spectrum against gauge fluctuations}
\label{dis}
 
As shown in the previous Sections, a translationally invariant 
non-Abelian potential does not change the size of the MBZ. 
This fact has important consequences on the stability of the 
Weyl semimetal phase to non-Abelian gauge fluctuations, such as 
variations of $q_{x,y}$ in Eq.~\eqref{potrashba}.
 
In the presence of a purely non-Abelian coupling, obtained by 
imposing $m=0$ in Eq.~\eqref{potabgen}, all the operators $U$ loose their 
space-dependence and the MBZ coincides with the usual BZ induced 
by the geometric shape of the lattice. Therefore, no qualitative 
deviation from the spectrum at $L \to \infty$ is expected in the 
presence of small perturbations of the coefficients $q_i$. 
Moreover, the spectrum changes continuously with the strength of the 
non-Abelian component and no fractal structure is found. 
For this reason we expect a substantial stability of the spectrum 
against small fluctuations $\delta q_{x,y}$ of the non-Abelian gauge potential.

Instead, when both an Abelian and a uniform non-Abelian couplings are 
involved, the MBZ size is the same as in the purely Abelian case, so that 
a fractal instability in the limit $L \to \infty$ occurs 
from the Abelian contribution only. 
This expectation can be directly probed directly for the 
system described by Eq.~\eqref{harper2D}, where a double-Weyl 
semimetal phase appears. These semimetals are protected by a 
$C_4$ symmetry \cite{bernevig12} and characterized by isolated band 
touching points between two bands. In the same points the dispersion 
is quadratic along two momentum directions and linear along the third one, 
so that they are characterized by a linearly vanishing density of states.

\begin{figure}[h!]
\centering
\includegraphics[width=1.0\textwidth]{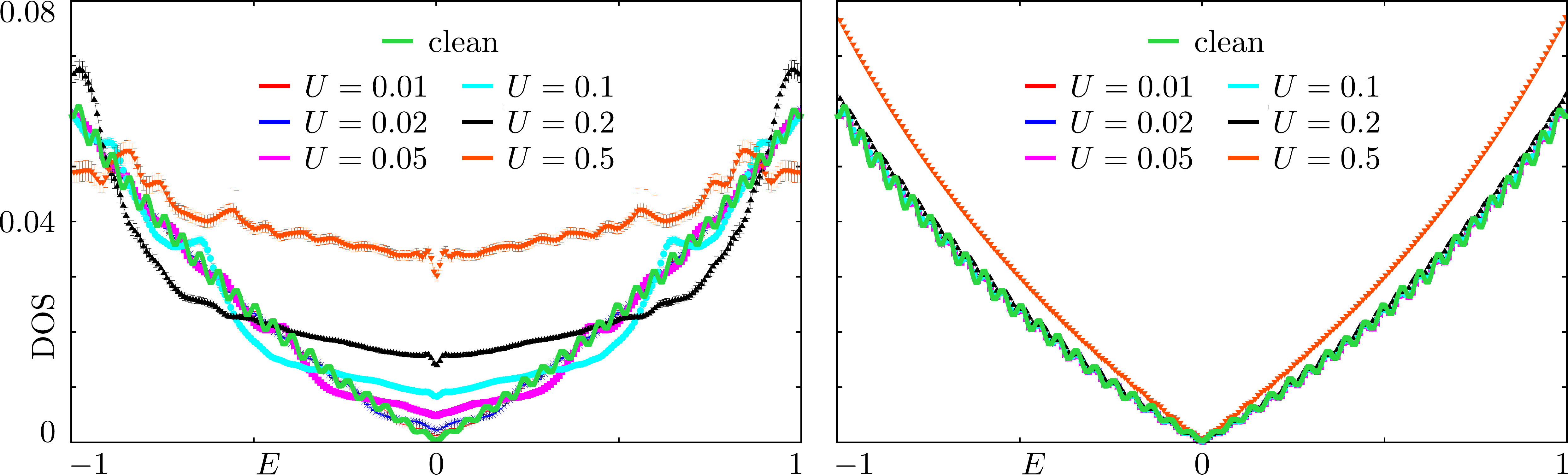}
\caption{Average density of states $\rho$ as a function of energy (in units of $t$)
around the double-Weyl points for random fluctuations of the flux 
$\Phi$ around $\pi$ (left panel) or $q_{x,y}$ around $\pi/4$ (right panel). 
Different colors denote the strength of the fluctuations, $U$. For $U=0$, 
$\rho$ is linear around the double-Weyl point energy ($E=0$) in both cases. 
In the left panel, a flat region of nonzero density of states 
develops even for small values of $U$, signaling the destruction 
of the semimetal phase. In the right panel, $\rho$ 
remains linear also for large values of $U$, such that the double-Weyl 
semimetal is robust to fluctuations in the non-Abelian potential only.}   
\label{twodis}
\end{figure} 

In Fig.~\ref{twodis} the stability of the band-touching points 
against fluctuations of the Abelian flux $\delta \Phi$ and 
of the non-Abelian strengths $\delta q_{x,y}$ is investigated 
(left and right panel, respectively), for a cubic system of 
linear size $L=120$. The fluctuations are randomly and independently 
drawn from the uniform distribution $[-U, U]$ around the values 
$\Phi=\pi$ and $q_x=q_x=\pi/4$. We show the density of states 
at energies around the band touching points, calculated by 
averaging over $200$ independent random potential configurations 
at fixed disorder strength $U$, using a kernel polynomial approximation 
with $N=1024$ polynomials, and a stochastic evaluation of the trace with 
$R=10$ random vectors (for details on the method, see Ref.~\cite{lepori2015}). 
In the absence of fluctuations, the density of states 
shows the expected linear profile around the band touching points.
 
At fixed $q_x=q_y=\pi/4$ and adding fluctuations in $\Phi$, we find that the 
density of states at the energy of the double Weyl points 
rapidly develops a plateau as $U$ increases, signaling the breakdown 
of the double-Weyl semimetal. The situation in very similar a
to the one described in Ref.~\cite{lepori2016} for the purely Abelian 
$\pi$ flux cubic lattice model. At variance, 
at fixed $\Phi = \pi$ and for fluctuating $q_{x,y}$ the density of states 
remains linear also for larger values of $U$. These results support 
our expectation on the stability of the spectra 
against fluctuation of the gauge potentials, based on the discussion on the 
properties of the MBZ presented above.  
 
\section{Conclusions}

In this work we presented the general analytic formalism 
to solve tight-binding models describing particles 
in a cubic lattice subject to translational invariant Abelian and 
non-Abelian gauge potentials. We considered the general case of 
commensurate magnetic fluxes, possibly different along the three directions.

We then discussed several examples related to $U(2)$ potentials, 
 as systems with a Rashba-like coupling, also relevant for 
the realization of particular topological semimetals \cite{lepori2016}.
There we also investigated the effects of perturbing the magnetic fluxes.

Our study illustrates the interexchange between the formal techniques 
developed to describe lattice particles in magnetic fields 
and the advancements coming from the simulation 
of tunable gauge potentials in ultracold atom experiments. These adavancements
motivate the exploration of the mathematical structure of the single particle 
energy spectrum in new situations, such as in the presence of non-Abelian gauge 
potentials. 

We showed in particular that the Hasegawa gauge, in the presence 
of a commensurate Abelian flux $\Phi=2 \pi \frac{m}{n}$, 
reduces the problem in the momentum space to the diagonalization 
of a $n \times n$ matrix for the purely Abelian case, and of a 
$pn \times pn$ matrix if the particle has 
$p$ degrees of freedom (more generally of the system has $p$ components). 
Exploiting this formalism, one can study the case of vanishing 
flux $\Phi \to 0$ (i.e., $n \to \infty$) or the dependence of the energy 
spectrum on the parameters of the translational invariant non-Abelian gauge 
potential, such as $q$ for the potential $\vec{A}_{\text{NAB}} = 
q \, (\sigma_x , \sigma_y, \sigma_z)$, that we considered in Section 
\ref{examplestot}.

Our study also offers useful tools for the design and analysis of 
ultracold atom and photonic platforms for the realization of exotic 
topological phases of matter, often relying on artificial gauge potentials, 
and it provides interesting alternative routes for the implementation of 
novel quantum phenomena. 
On the technical level, the approach described here can be generalized 
also to the case of magnetic fluxes which vary periodically across the lattice 
and it can be also extended to the simulations of the so-called extra dimensions  \cite{extra}, a topic recently 
at the center of many discussions in the quantum simulation 
community, for instance concerning the experimental realization of 4D quantum Hall systems 
\cite{price15,lohse17}. 

\section*{Acknowledgements}
The authors thank A. Celi, L. Fallani, G. Juzeli{\=u}nas, M. Mannarelli,  
and S. Paganelli for useful discussions. 
M.B., L.L. and A.T. thank the Galileo Galilei 
Institute for Theoretical Physics, Firenze, for the hospitality in the
Workshop ``From Static to Dynamical Gauge Fields with
Ultracold Atoms'', 22th May - 23th June 2017, and the INFN for partial support during  the  completion of this  work. M.B. acknowledges Villum Foundation for support.

\end{document}